\documentclass[lettersize,journal]{IEEEtran}
\usepackage{amsmath,amsfonts}
\usepackage{amsthm}
\usepackage{array}
\usepackage[caption=false,font=normalsize,labelfont=sf,textfont=sf]{subfig}
\usepackage{textcomp}
\usepackage{stfloats}
\usepackage{url}
\usepackage{verbatim}
\usepackage{graphicx}
\usepackage{cite}
\usepackage{subfig}
 \usepackage{threeparttable}
 \usepackage{makecell}
\usepackage[table]{xcolor} 
\makeatletter
\newif\if@restonecol
\makeatother

\usepackage[linesnumbered,ruled,vlined]{algorithm2e}
\usepackage{algpseudocode}

\begin{document}
 
\title{Joint Computation Offloading and Resource Management for Cooperative Satellite-Aerial-Marine Internet of Things Networks}

\author{
Shuang Qi,~\IEEEmembership{Student Member,~IEEE,}
Bin Lin,~\IEEEmembership{Senior Member,~IEEE,}
Yiqin Deng,
Hongyang Pan,
Xu Hu,~\IEEEmembership{Student Member,~IEEE}\vspace{-2em}
\thanks{The work of Bin Lin was supported in part by the National Natural Science Foundation of China (No. 62371085) and in part by the Fundamental Research Funds for the Central Universities (No. 3132023514). The work of Yiqin Deng was supported in part by the National Natural Science Foundation of China (No. 62301300). \textit{(Corresponding author: Bin Lin.)}}
\thanks{Shuang Qi, Bin Lin, Hongyang Pan and Xu Hu are with the Information Science and Technology College, Dalian Maritime University, Dalian 116026, China (e-mail: qishuang\_0315@163.com; binlin@dlmu.edu.cn; panhongyang18@foxmail.com; huxu@dlmu.edu.cn).}
\thanks{Yiqin Deng is with Hong Kong JC STEM Lab of Smart City and Department of Computer Science, City University of Hong Kong, Kowloon, Hong Kong, China (email: yiqideng@cityu.edu.hk).}
\thanks{This manuscript has been accepted by IEEE Internet of Things Journal, DOI: 10.1109/JIOT.2025.3617096.}}
\maketitle

\begin{abstract}
Devices within the marine Internet of Things (MIoT) can connect to low Earth orbit (LEO) satellites and unmanned aerial vehicles (UAVs) to facilitate low-latency data transmission and execution, as well as enhanced-capacity data storage. However, without proper traffic handling strategy, it is still difficult to effectively meet the low-latency requirements. In this paper, we consider a cooperative satellite-aerial-MIoT network (CSAMN) for maritime edge computing and maritime data storage to prioritize delay-sensitive (DS) tasks by employing mobile edge computing, while handling delay-tolerant (DT) tasks via the store-carry-forward method. Considering the delay constraints of DS tasks, we formulate a constrained joint optimization problem of maximizing satellite-collected data volume while minimizing system energy consumption by controlling four interdependent variables, including the transmit power of UAVs for DS tasks, the start time of DT tasks, computing resource allocation, and offloading ratio. To solve this non-convex and non-linear problem, we propose a joint computation offloading and resource management (JCORM) algorithm using the Dinkelbach method and linear programming. Our results show that the volume of data collected by the proposed JCORM algorithm can be increased by up to 41.5\% compared to baselines. Moreover, JCORM algorithm achieves a dramatic reduction in computational time, from a maximum of 318.21 seconds down to just 0.16 seconds per experiment, making it highly suitable for real-time maritime applications.
\end{abstract}

\begin{IEEEkeywords}
marine Internet of Things (MIoT), computation offloading, unmanned aerial vehicles (UAVs), mobile edge computing, resource allocation
\end{IEEEkeywords}

\section{Introduction}
\IEEEPARstart{T}{he} impending advancement of the sixth-generation non-terrestrial networks (NTN) for aerospace vehicles significantly propels the development of the marine Internet of Things (MIoT) system. This system leverages the seamless coverage advantages of satellites in space and the flexible accessibility of unmanned aerial vehicles (UAVs) in the air, providing solutions for massive data collection, transmission, and processing in MIoT \cite{9184929} \cite{Chen_2024}. Massive marine data, including text, voice, images, and video, play a vital role in multiple scenarios, including marine environmental monitoring, maritime search and rescue, and natural disaster recovery \cite{Hu2024Performance}. A fundamental characteristic of MIoT applications is the diversity of service requirements coexisting within the same temporal window. These disparate tasks can be broadly categorized into delay-sensitive (DS) tasks, where low latency is paramount \cite{10118864} \cite{10130060}, and delay-tolerant (DT) tasks, where objectives shift towards maximizing data volume while managing energy consumption and operational costs \cite{9759502} \cite{9701330} \cite{9611538}. To effectively serve both task types, designing distinct processing modes to their individual needs is essential for enhancing the overall resource utilization efficiency of the MIoT.

To reduce the execution delay of DS tasks, integrating mobile edge computing (MEC) servers on UAVs and low Earth orbit (LEO) satellites is an effective solution \cite{10804104}. By performing data computation and analysis at the edge of the network \cite{9685920}, only the results are transmitted back to the shore-based control center, thus avoiding significant transmission delays. Considering the limited computational capabilities and energy of UAVs, employing computation offloading techniques allows UAVs to transmit data for processing onto the LEO satellite, thus meeting the delay requirements for handling DS tasks \cite{10278101}. However, although excessive offloading is beneficial to the computation speed, it intensifies the contention for communication resources and the energy consumption of UAVs, while insufficient offloading can cause the onboard processor of UAVs to overload. Finding the right offloading ratio is therefore a critical problem. 

For DT tasks, the use of store-and-forward (SCF) technology is an effective solution. By temporarily storing DT data in the storage spaces of UAVs and LEO satellites, DT data transmission is only carried out after the completion of DS tasks, thereby avoiding conflicts with DS tasks. On the one hand, this approach can prevent occupying the bandwidth of backhaul links, which could affect the transmission rate of DS data. On the other hand, it avoids competing for network resources with DS tasks. In this process, it is necessary to consider how to allocate network resources for DS tasks and DT tasks to meet the performance requirements of both types of tasks simultaneously.

Motivated by the above analysis, existing studies typically focus separately on optimizing either DS task offloading/processing or DT task scheduling/collection within homogeneous networks. However, in practical scenarios, DS and DT tasks coexist and compete for limited communication and computing resources. The joint optimization of computation offloading and cross-layer resource management in heterogeneous satellite-aerial networks supporting heterogeneous services has not been fully explored. There is a pressing need for a unified framework to coordinate these interdependent factors to maximize the system utility while satisfying the different quality-of-service (QoS) requirements of both DS and DT tasks. Thus, this paper aims to simultaneously deal with DS and DT tasks in the cooperative satellite-aerial-MIoT network (CSAMN). The main contributions are summarized as follows.

\begin{itemize}
 \item In the scenario where both DS and DT tasks coexist, we present the CSAMN architecture combined two data processing modes, including the LEO satellite and UAV collaborative edge computing mode for DS data processing, and the LEO satellite SCF mode for DT data collection.
 \item Considering the delay requirement of DS tasks, we propose a joint computation offloading and resource management (JCORM) scheme designed to maximize utility, thereby achieving simultaneous maximization of the total data volume collected by LEO satellite and minimization of the total energy consumption of the CSAMN. We jointly optimize the transmit power of UAVs for DS tasks, the start time of DT tasks, computing resource allocation, and offloading ratio.
 \item The JCORM algorithm is proposed to solve the formulated non-convex and non-linear problem through alternating iterative optimization. The numerical results show that the proposed JCORM algorithm outperforms the benchmark methods in both utility maximization and computational efficiency, demonstrating rapid convergence characteristics.
\end{itemize}

The remainder of this paper is organized as follows. Section II introduces the related works. In Section III, we present the system model. In Section IV, we formulate the problem formulation. In Sections V and VI, the proposed algorithm and its evaluation are given, respectively. Finally, Section VII concludes this paper.

\section{Related Works}
In this section, we discuss the works related to computation offloading and resource allocation in space-air-ground integrated networks (SAGINs). Then, we review the work related to the space-air-ground-sea integrated networks (SAGSINs). Finally, we analyze the research gap.
\subsection{SAGINs}
SAGIN, as an integration of satellite systems, air networks and terrestrial communications, has become an emerging architecture and has attracted intense research interest over the past years \cite{8368236}. Masayuki \textit{et al.} \cite{10599115} explained the challenges of the optical-based SAGIN and proposed the machine learning and smart networking solutions. Chen \textit{et al.} \cite{11039742} discussed the key challenges and promising technologies in SAGIN for emergency communication systems. As user demands evolve, SAGIN with computational capabilities emerges as a viable solution for services that require low latency and high computational power. Xiao \textit{et al.} \cite{10745905} offered a comprehensive survey of the technological advances in communication and computing within SAGIN for the sixth generation. Chen \textit{et al.} \cite{10579820} offered a detailed overview of resource management modeling and optimization methods in joint communication and computing-embedded SAGIN, encompassing both traditional optimization approaches and learning-based intelligent decision-making frameworks. In summary, SAGINs have been further considered as a promising technical paradigm for global coverage and computing platforms.

In SAGIN, some studies employ UAVs to provide MEC services while using LEO satellites as relay nodes to enable ubiquitous cloud computing access. Jaiswal \textit{et al.} \cite{Jaiswal_2022} optimized the offloading decisions of users to maximize the number of computational tasks successfully processed under energy and delay constraints. Jiang \textit{et al.} \cite{10978788} minimized energy consumption by jointly optimizing offloading decisions, task scheduling, and UAV trajectory. Nguyen \textit{et al.} \cite{Nguyen_2022} used LEO satellites to offload tasks to cloud servers, aiming to minimize energy consumption while meeting delay constraints. Mao \textit{et al.} \cite{Mao_2020} optimized computation offloading and resource allocation, achieving a 30\% reduction in maximum computational delay compared to air-ground cooperative schemes. However, the aforementioned studies are limited to the communication relay function of LEO satellites, failing to fully exploit the potential advantages of on-board MEC.

To fully use LEO satellites, deploying MEC servers on LEO satellites enables UAVs to offload tasks onto satellites, thereby improving the efficiency of task processing. The computation offloading is fundamentally a distributed optimization problem. Chen \textit{et al.} \cite{10689514} formulated the offloading decision problem as a game theory model, aiming to minimize the cost of all devices under the constraint of the visible time window of the satellite. Zhang \textit{et al.} \cite{10804104} proposed a stochastic optimization problem to minimize the time-average expected service delay by jointly optimizing resource allocation and task offloading while satisfying energy constraints. Diallo \textit{et al.} \cite{10978672} defined a system cost function that includes both energy consumption and task dropping cost, and formulated a constrained system cost minimization problem by jointly optimizing power allocation, task offloading and scheduling and UAV trajectory. Li \textit{et al.} \cite{10891825} formulated a joint optimization problem including UAV trajectory, task offloading, and bandwidth allocations, aiming to maximize long-term energy efficiency. All of the above studies adopt the binary offloading. In partial offloading, the offloading location and offloading ratio are usually jointly optimized. Pervez \textit{et al.} \cite{pervez2025efficient} proposed a queue-aware efficiency maximization problem by jointly optimizing task splitting and offloading, server selection, transmit power, path control and computation resource allocation.
\subsection{SAGSINs}
By integrating LEO satellites, UAVs, and buoys/ships among other heterogeneous nodes, SAGSINs are established to deliver low-latency, high-reliability communication and computing services for maritime users. Wang \textit{et al.} \cite{9453860} established a hierarchical framework for enhancing energy-efficient maritime coverage, using satellites for global signaling, offshore BS for sending source data, UAVs for on-demand filling of coverage blind spots, and vessels for opportunistic relaying data. Wang \textit{et al.} \cite{9869801} proposed a SAGSIN architecture that supports the sixth generation to achieve seamless coverage of global wireless signals.

Considering the accessibility of LEO satellites, Jung \textit{et al.} \cite{Jung_2023} proposed three solutions: always-on, always-off, and the intermediate disconnection. By jointly optimizing communication and computing resources, these solutions aim to minimize the energy consumption of UAVs. To minimize energy consumption under latency constraints, Lin \textit{et al.} \cite{10518032} optimized offloading strategy and bandwidth allocation. Wang \textit{et al.} \cite{10966051} jointly optimized offloading mode, offloading ratio, and computing resource allocation. Wu \textit{et al.} \cite{10912482} considered the computing offloading assisted by multiple high altitude platforms, aiming to minimize the total delay by jointly optimizing task scheduling, computing resource allocation, and task-satellite association.
\subsection{Summary}
All the above studies use UAVs and satellites to assist data transmission and processing. Nevertheless, the type variations are not considered in practical application scenarios. For different data types of DT data and DS data, different transmission modes can be considered for corresponding performance gains. Jia \textit{et al.} \cite{9184929} proposed two transmission modes, i.e., a UAV SCF mode for DT data and a satellite network relay mode for DS data transmission. Yao \textit{et al.} \cite{10194584} proposed the data collection efficiency maximization problem by optimizing UAV deployment, bandwidth allocation and transmit power. In their approach, however, DS data were transmitted in real time to the data center via LEO satellite, while UAVs stored and carried DT data until the acquisition was complete. Compared with the random scheme, the data collection efficiency growth rate of the scheme proposed in \cite{10194584} is approximately 133\%.

However, there are still some major issues remained. Firstly, UAVs combined with SCF technology for collecting DT data ignores the limited storage capacity of UAVs and energy consumption of multiple frequent round trips between collection points and the data center. Secondly, only the transmission mode of DS data is considered, while largely ignoring the computing issues (e.g., MEC). Finally, simultaneously handling DS data and DT data inevitably affects the efficiency of processing DS data. How to meet the QoS requirements of both tasks is a significant challenge. Therefore, this paper presents two data processing modes in CSAMN, including the LEO satellite and UAV collaborative edge computing mode for DS data processing and the LEO satellite SCF mode for DT data collection. We fully use the dual role of LEO satellites as storage units and MEC servers to support UAV operations, thereby further reducing DS data processing latency. Furthermore, the start time of the DT task is optimized to ensure the DS data is processed preferentially in CSAMN, thus guaranteeing compliance with its delay requirements.
\section{System Models}
First, we introduce the CSAMN architecture and tasks processing modes. Then, we present the communication models, including MIoT device-UAV communication model and UAV-LEO satellite communication model. The computing model for DS tasks and storage models for DT tasks are presented in Section III-C and Section III-D, respectively. Finally, we present the energy consumption models of CSAMN.
\begin{figure}[t]
\centering 
{\includegraphics[width=0.5\textwidth]{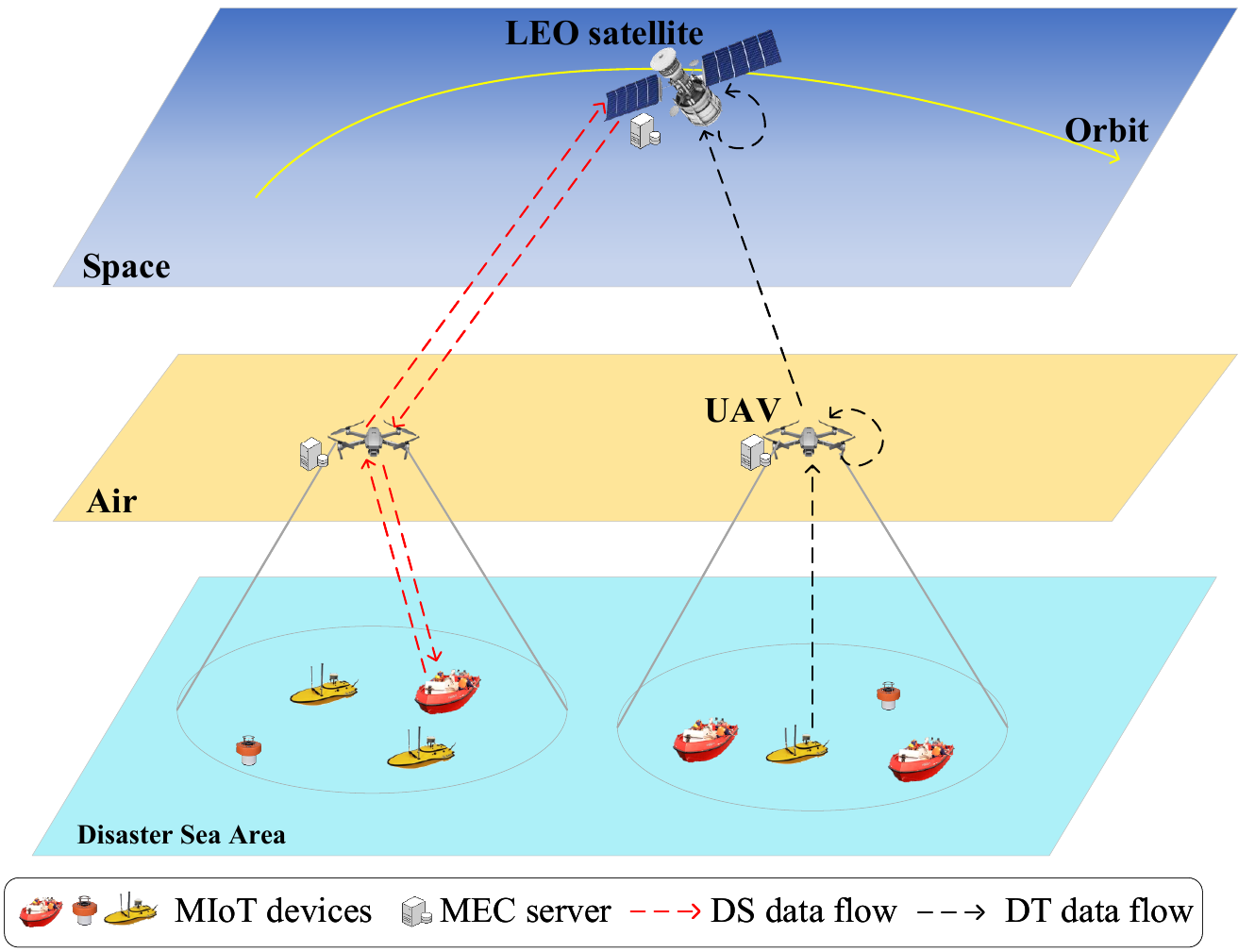}}
\caption{CSAMN architecture.}
\label{Fig.1}
\end{figure}

\subsection{Network Model}
\begin{figure}[t]
\centering 
{\includegraphics[width=0.45\textwidth]{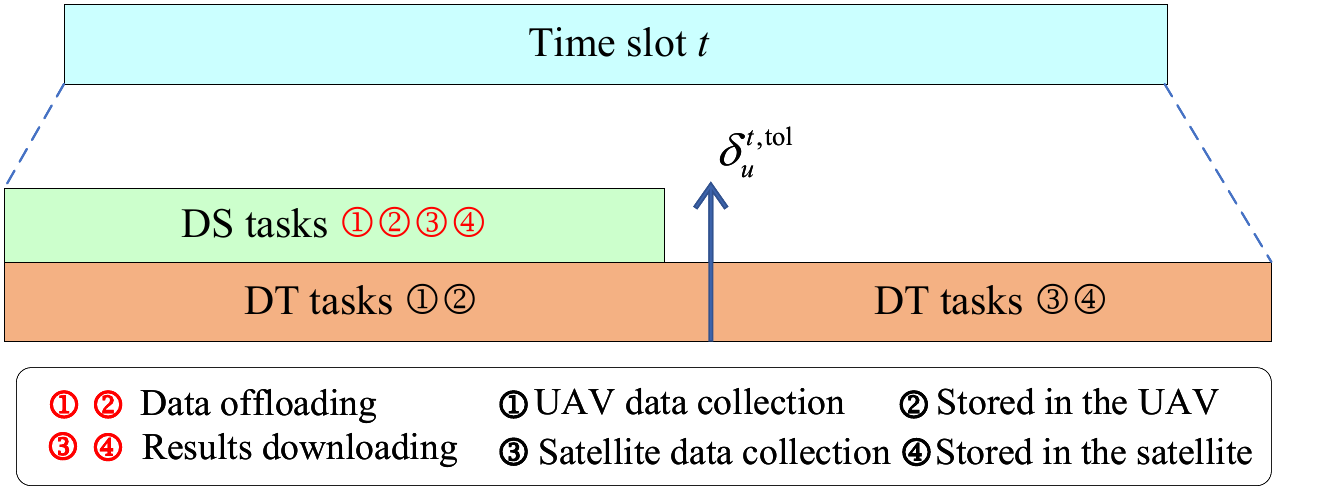}}
\caption{Tasks processing within a single time slot.}
\label{Fig.2}
\end{figure}

Fig. \ref{Fig.1} illustrates the CSAMN architecture, comprising multiple MIoT devices, multiple UAVs, and an LEO satellite. The set of UAVs is denoted as $\mathcal{U}=\{1,2,...,U\}$, which are uniformly distributed across the disaster-affected marine region, providing computational support, data collection, and store-and-forward services. Disaster-affected marine region can refer to areas such as those for post-shipwreck search and rescue or environmental monitoring during oil spills. Within the coverage area of each UAV, multiple DS maritime devices (DS-MDs) and DT maritime devices (DT-MDs) are randomly distributed. It is important to note that the classification of a device as a DS-MD or DT-MD is based on the latency requirement of the specific task it is currently handling, not its inherent hardware capabilities. A single device (e.g., a rescue ship) can function as both a DS-MD and a DT-MD at different times, depending on the application. The sets of DS-MDs and DT-MDs covered by UAV $u$ are denoted as $\mathcal{K}_u^{\text{sens}}=\{1,2,...,K_u^{\text{sens}}\}$ and $\mathcal{K}_u^{\text{tol}}=\{1,2,...,K_u^{\text{tol}}\}$, respectively. The visible time window for the LEO satellite is defined as 
\begin{equation}
T_{\text{v}}=\frac{2(r_{\text{E}}+h)\gamma_{\text{LEO}}}{v_{\text{s}}},
\end{equation}
where $h$ is the altitude of LEO satellite, $r_\text{E}$ is the radius of the Earth, and $v_\text{s}$ is the speed of the LEO satellite. The angle $\gamma_\text{LEO}$ of the satellite coverage is calculated by
\begin{equation}
\gamma_{\text{LEO}}=\arccos{\Big(\frac{r_{\text{E}}}{r_{\text{E}}+h}\cdot\cos\theta\Big)}-\theta,
\end{equation}
where $\theta$ is the elevation angle of the satellite\cite{Jung_2023}\cite{9344666}. The task processing time is segmented into $T$ time slots, and the set of time slots is denoted by $\mathcal{T}=\{1,2,...,T\}$. Each time slot, denoted as $t\in\mathcal{T}$, has a duration of $\delta$, which satisfies the condition $\delta T\leq T_v$. Furthermore, each time slot is split into two parts to provide services for DS tasks and DT tasks separately. Fig. \ref{Fig.2} presents the tasks processing within a single time slot. The foremost priority in each time slot is the immediate servicing of all DS tasks. Any DT tasks are deferred. DT task processing is permitted to initiate flexibly at any time during the resulting residual period once the high-priority workload is cleared. Let $\delta_{u}^{t,\text{tol}}\in[0,\delta]$ represents the start time of the DT task at the time slot $t$. Specifically, we elaborate on the processing procedure for these two task categories within each time slot. 
\begin{itemize}
\item {\it{DS tasks processing}}: Initially, DS-MD $k_u^\text{sens}\in\mathcal{K}_u^{\text{sens}}$ assigns the entire computing task to UAV $u$. Subsequently, UAV $u$ forwards the computing task to the LEO satellite for cooperative processing through partial offloading. Thereafter, the LEO satellite executes the computing task, and the results are transmitted back to UAV $u$. Finally, the results from both the LEO satellite and UAV $u$ are returned to the DS-MD $k_u^\text{sens}$.

\item {\it{DT tasks processing}}: In the proposed CSAMN architecture, the initial two steps of the DT tasks proceed concurrently with the DS tasks. Specifically, the UAV collects data from DT-MDs and stores it locally. Subsequent to completion of the DS task processing, the UAV forwards the DT data to the LEO satellite. Ultimately, leveraging SCF technology, the LEO satellite stores the data and transmits it back to a shore-based data center when positioned above the ground station. This approach effectively alleviates the burden on the backhaul link. 
\end{itemize}

\subsection{Communication Models}
In the CSAMN framework, transmission links include links among MIoT devices and UAVs, as well as links among UAVs and the LEO satellite. 
\subsubsection{MIoT Device-UAV Communication Model}
Let the location of the UAV at time slot $t$ is $\mathbf{q}_{u}^{t}=(x_{u}^{t}, y_{u}^{t}, h_{u})$. We assume that the altitude $h_{u}$ of the UAV $u$ is unchanged in each time slot. Let MIoT device $k$ is deployed at position $\mathbf{q}_{k}^{t}=(x_{k}^{t}, y_{k}^{t},0)$ in time slot $t$, where $k\in\mathcal{K}_u^\text{sens}\cup \mathcal{K}_u^\text{tol}$. Considering the uniqueness of the marine environment, e.g., the strong direct signal, the primary factors affecting the overseas wireless channel are multi-path effects caused by ocean waves and weather conditions \cite{10966051}. The MIoT device-to-UAV link is modeled as a combination of distance-dependent large-scale path loss and small-scale Rician fading \cite{11007596}.

The large-scale path loss model is expressed as
\begin{equation}
G_{k,u}^{t}=PL_\text{c}(d_{k,u}^{t})^{-PL_\text{e}},
 \quad \forall k,u,t,
\end{equation}
where  $d_{k,u}^{t}$,  $PL_\text{c}$, and $PL_\text{e}$ are the distance between the MIoT device $k$ and UAV $u$, the path loss coefficient, and the path loss exponent for MIoT device-to-UAV link, respectively \cite{11007596}.

The small-scale Rician fading is represented as
\begin{equation}
\hat{G}_{k,u}^{t}=\sqrt{\frac{K_0}{1+K_0}}+\sqrt{\frac{1}{1+K_0}}o_{k,u}^{t},
 \quad \forall k,u,t,
\end{equation}
where, $o_{k,u}^{t}\in\mathcal{CN}(0,1)$ and $K_0$ is the Rician factor \cite{10966051}. Then, the channel gain from MIoT device $k$ to the UAV $u$ is $g_{k,u}^{t}=G_{k,u}^{t}\hat{G}_{k,u}^{t}$. Assuming that DS-MDs and DT-MDs adopt mutually orthogonal sub-channels. Let $\beta_u^{t}\in[0,1]$ denotes the fraction of bandwidth allocated by the UAV to DS-MDs. Therefore, the achievable rate from DS-MD $k_{u}^\text{sens}$ to UAV $u$ is
\begin{equation}
R_{k_{u}^\text{sens},u}^{t}=\frac{\beta_u^{t}B_{u}}{K_{u}^\text{sens}}\log_2\Big(1+\frac{p_{k_{u}^\text{sens},u}^{t}g_{k_{u}^\text{sens},u}^{t}}{\sigma^{2}}\Big),\quad \forall k_{u}^\text{sens},u,t,
\end{equation}
and the achievable rate from DT-MD $k_{u}^\text{tol}$ to UAV $u$ is
\begin{equation}
R_{k_{u}^\text{tol},u}^{t}=\frac{(1-\beta_u^{t})B_{u}}{K_{u}^\text{tol}}\log_2\Big(1+\frac{p_{k_{u}^\text{tol},u}^{t}g_{k_{u}^\text{tol},u}^{t}}{\sigma^{2}}\Big),\quad \forall k_{u}^\text{tol},u,t,
\end{equation}
where $B_u$ is the available bandwidth and $\sigma^2$ is the noise power at the receiver. $p_{k_{u}^\text{sens},u}^{t}$ and $p_{k_{u}^\text{tol},u}^{t}$ are the transmit power of the DS-MD $k_{u}^\text{sens}$ and DT-MD $k_{u}^\text{tol}$, respectively.

\subsubsection{UAV-LEO Satellite Communication Model}
With the development of microwave communications and multibeam antenna technique, the feasibility of direct ground-space transmission has been demonstrated \cite{Mao_2020}.  Frequency division multiple access manner is employed for communication between the LEO satellite and UAVs at the time slot $t$. We assume that all UAVs have DS tasks requiring computation in each time slot. However, uploading DT data is not required in each time slot. The channel gain from UAV $u$ to the LEO satellite is
\begin{equation}
g_{u,\text{LEO}}^{t}=\frac{g_0G}{(d_{u,\text{LEO}}^{t})^{2}}=\frac{g_0G}{\Vert \mathbf{q}_{u}^{t}-\mathbf{q}_{\text{LEO}}^{t} \Vert^{2}}, \quad \forall u,t,
\end{equation}
where $G$ is an antenna gain for the long distance satellite communication consisting of the transmission antenna gain of the UAV and the receiver antenna gain of the LEO satellite. The distance between UAV $u$ and the LEO satellite can be calculated by \cite{Jung_2023} \cite{9344666}
\begin{equation}
d_{u,\text{LEO}}^{t}=\frac{(r_\text{E}+h)\sin{\gamma_{\text{LEO}}}}{\cos{\theta}}.
\end{equation}

Therefore, the achievable rate from UAV $u$ to the LEO satellite for DS tasks and DT tasks are
\begin{equation}
R_{u,\text{LEO}}^{t,\text{sens}}=\frac{B_\text{LEO}}{U}\log_2\Big(1+\frac{p_{u,\text{LEO}}^{t,\text{sens}}g_{u,\text{LEO}}^{t}}{\sigma^{2}}\Big),\quad \forall u,t,
\end{equation}
\begin{equation}
R_{u,\text{LEO}}^{t,\text{tol}}=\frac{B_\text{LEO}}{U}\log_2\Big(1+\frac{p_{u,\text{LEO}}^{t,\text{tol}}g_{u,\text{LEO}}^{t}}{\sigma^{2}}\Big),\quad \forall u,t,
\end{equation}
respectively. $B_\text{LEO}$ is the bandwidth. $p_{u,\text{LEO}}^{t,\text{sens}}$ and $p_{u,\text{LEO}}^{t,\text{tol}}$ are the transmit power of UAV $u$ for DS tasks and DT tasks, respectively.

\subsection{Computing Model for DS Tasks}
For DS tasks, the partial offloading mode is employed by the UAV. Let the binary variable $\gamma_{u,\text{LEO}}^{t}\in[0,1]$ denotes the offloading ratio of  UAV $u$ to the LEO satellite. The data size of the DS-MD $k_{u}^{\text{sens}}$ in time slot $t$ is $D_{k_{u}^{\text{sens}}}^{t}$ and the computing capability of UAV $u$ is $F_u$. According to the above processing procedure of DS tasks, we present the time cost of each step in the following.

In the first step, the DS-MD $k_{u}^\text{sens}$ offloads the entire computing task to UAV $u$. To ensure that the task transmission from each DS-MD is completed, the time cost of DS tasks transmission from DS-MDs to UAV $u$ can be represented as
\begin{equation}
l_{u}^{t,\text{off}}=\max\Big\{\frac{D_{k_{u}^{\text{sens}}}^{t}}{R_{k_{u}^{\text{sens}},u}^{t}}, \forall k_{u}^{\text{sens}}\in K_{u}^{\text{sens}}\Big\}, \quad \forall u,t.
\end{equation}

In the second step, UAV $u$ executes DS tasks and transmits the tasks to the LEO satellite cooperative processing by partial offloading. The time cost consists of the local computing time and data offloading time from UAV $u$ to the LEO satellite at time slot $t$. The local computing time is represented as 
\begin{equation}
l_{u}^{t,\text{comp}}=\frac{f_0(1-\gamma_{u,\text{LEO}}^{t})\sum_{k_{u}^{\text{sens}}=1}^{K_u^{\text{sens}}}D_{k_{u}^{\text{sens}}}^{t}}{F_u}, \quad \forall u,t,
\end{equation}
where $f_0$ is the CPU cycles required to process one bit data.

Due to the large physical distance, the delay of task offloading to LEO satellite includes transmission delay, propagation delay, and computing delay \cite{9861225}. The transmission delay can be calculated by
\begin{equation}
l_{u,\text{LEO}}^{t,\text{comm}}=\frac{\gamma_{u,\text{LEO}}^{t}\sum_{k_{u}^{\text{sens}}=1}^{K_u^{\text{sens}}}D_{k_{u}^{\text{sens}}}^{t}}{R_{u,\text{LEO}}^{t,\text{sens}}}, \quad \forall u,t.
\end{equation}
The propagation delay can be calculated by
\begin{equation}
l_{\text{LEO}}^{t,\text{prop}}=d_{u,\text{LEO}}^{t}/{c}, \quad \forall u,t,
\end{equation}
where $c$ is the speed of light. 

Let the computing capability of the LEO satellite be $F_\text{LEO}$ and the computing resources allocated by the LEO satellite to UAV $u$ at time slot $t$ be $f_{u,\text{LEO}}^{t}$. Then, the time cost of task execution in the LEO satellite is represented as
\begin{equation}
l_{u,\text{LEO}}^{t,\text{comp}}=\frac{f_0\gamma_{u,\text{LEO}}^{t}\sum_{k_{u}^{\text{sens}}=1}^{K_u^{\text{sens}}}D_{k_{u}^{\text{sens}}}^{t}}{f_{u,\text{LEO}}^{t}}, \quad \forall u,t.
\end{equation}

Finally, since the data size of the execution result is typically very small, the transmission delay for returning the results can be reasonably neglected. However, considering the altitude of the LEO satellite, the propagation delay of the results should be considered. When the UAV receives the computing task, it simultaneously executes the task locally and offloads it. Therefore, the total time cost of UAV $u$ for DS tasks processing is expressed as
\begin{equation}
l_{u}^{t}=l_{u}^{t,\text{off}}+\max\{l_{u}^{t,\text{comp}},l_{u,\text{LEO}}^{t,\text{comm}}+l_{u,\text{LEO}}^{t,\text{comp}}+2l_{u,\text{LEO}}^{t,\text{prop}}\}, \quad \forall u,t.
\end{equation}

\subsection{Storage Models for DT Tasks}
\subsubsection{DT Data Collection from DT-MDs to UAVs}
During the processing of DS tasks, the UAV continuously collects maritime data and stores them in the storage space. At time slot $t$, the data volume from DT-MDs collected by UAV $u$ is given as follows:
\begin{equation}
D_u^t=\sum_{k_{u}^{\text{tol}}=1}^{K_u^{\text{tol}}}D_{k_{u}^{\text{tol}},u}^t=\sum_{k_{u}^{\text{tol}}=1}^{K_u^{\text{tol}}}R_{k_{u}^\text{tol},u}^{t}\delta_{u}^{t,\text{tol}},\quad \forall k_{u}^{\text{tol}},u,t.
\end{equation}
Assuming that the storage capability and the initial remaining storage space of UAV $u$ are $D_u^{\text{stor}}$ and $D_u^{0,\text{stor}}$, respectively. At the time slot $t$, the data collected by UAV $u$ cannot exceed its remaining storage space, i.e.,
\begin{equation}
D_u^t\leq D_{u}^{t,\text{stor}}, \quad \forall u,t.
\end{equation}

In this paper, we assume that the DT-MDs continuously collect data. Therefore, the effect of the data volume in DS-MDs on the data volume collected by the UAV is ignored.

\subsubsection{DT Data Collection from UAVs to LEO Satellite}
We assume that the storage capacity of the LEO satellite is large enough. The data volume that the LEO satellite is able to collect from UAV $u$ during the data collection period is denoted by
\begin{equation}
D_{u,\text{LEO}}^{t}=R_{u,\text{LEO}}^{t,\text{tol}}(\delta-\delta_u^{t,\text{tol}}), \quad \forall u,t.
\end{equation}

However, the data volume that LEO satellite can collect from a UAV is naturally limited by the data volume actually stored on that UAV, i.e,
\begin{equation}
D_{u,\text{LEO}}^{t}\leq D_{u,\text{LEO}}^{t,\text{max}}, \quad \forall u,t,
\end{equation}
where $D_{u,\text{LEO}}^{t,\text{max}}$ indicates the data volume stored by UAV $u$ at the time slot $t$. In other words, $D_{u,\text{LEO}}^{t,\text{max}}$ is the maximum data volume that can be uploaded to the LEO satellite, i.e.,
\begin{equation}
D_{u,\text{LEO}}^{t,\text{max}}=D_u^t+(D_{u}^{\text{stor}}-D_{u}^{t,\text{stor}}),  \quad  \forall  u,t.
\end{equation}

Note that if the data stored by the UAV is not fully collected by the LEO satellite, it continues to be stored in the UAV for subsequent transmission, and its storage capacity is also affected. Therefore, the remaining storage space of the UAV at time slot $t+1$ can be updated using the following formula: 
\begin{equation}
D_{u}^{t+1,\text{stor}}=\min\{D_{u}^{t,\text{stor}}-D_{u}^{t}+D_{u,\text{LEO}}^{t},D_{u}^{\text{stor}}\}, \quad  \forall  u,t.
\end{equation}

\subsection{Energy Consumption Models}
In energy consumption models, we consider both UAV energy consumption and LEO satellite energy consumption.
\subsubsection{UAV Energy Consumption Model}
Since the hovering time of UAVs is fixed in the given time, we only consider the communication and computing energy consumption of the UAV. The communication energy consumption of UAV $u$ can be expressed as
\begin{equation}
\begin{aligned}
E_{u,\text{LEO}}^{t,\text{comm}}=p_{u,\text{LEO}}^{t,\text{sens}}l_{u,\text{LEO}}^{t,\text{comm}}+p_{u,\text{LEO}}^{t,\text{tol}}(\delta-\delta_u^{t,\text{tol}}), \quad \forall u,t,
\end{aligned}
\end{equation}
where the first term and second term are the communication energy consumption for DS tasks and DT tasks, respectively. The computing energy consumption of UAV $u$ can be expressed as
\begin{equation}
E_{u}^{t,\text{comp}}=f_0\kappa(1-\gamma_{u,\text{LEO}}^{t})F_{u}^2\sum_{k_{u}^{\text{sens}}=1}^{K_u^{\text{sens}}}D_{k_{u}^{\text{sens}}}^{t}, \quad \forall u,t,
\end{equation}
where $\kappa$ denotes the effective switched capacitance of CPU, and its value is relevant to the clip architecture \cite{10490255}. Therefore, the total energy consumption of UAV $u$ at time slot $t$ is expressed as
\begin{equation}
E_{u}^{t}=E_{u,\text{LEO}}^{t,\text{comm}}+E_{u}^{t,\text{comp}},  \quad  \forall  u,t.
\end{equation}

\subsubsection{LEO Satellite Energy Consumption Model}
Since the data volume of results is very small, the communication energy consumption returned by the result is ignored. We only consider the computing energy consumption of the LEO satellite, which can be shown as
\begin{equation}
\begin{aligned}
E_{u,\text{LEO}}^{t,\text{comp}}=f_0\kappa\gamma_{u,\text{LEO}}^{t}(f_{u,\text{LEO}}^{t})^2\sum_{k_{u}^{\text{sens}}=1}^{K_u^{\text{sens}}}D_{k_{u}^{\text{sens}}}^{t}, \quad  \forall  u,t.
\end{aligned}
\end{equation}

In summary, the total energy consumption of the CSAMN at the time slot $t$ is represented as
\begin{equation}
\begin{aligned}
E^{t}=\sum_{u=1}^{U}(E_{u,\text{LEO}}^{t,\text{comm}}+E_{u}^{t,\text{comp}}+E_{u,\text{LEO}}^{t,\text{comp}}), \quad  \forall  t.
\end{aligned}
\end{equation}

\section{Problem Formulation}
To capture the heterogeneous QoS requirements of DS and DT tasks, this paper models the low latency requirement of DS tasks as constraints, while placing the trade-off between data collection performance of DT tasks and energy consumption within the objective function. We define the utility function as the difference between the data volume collected by the LEO satellite and the total energy consumption of the CSAMN. To enhance the utility of the CSAMN, the transmit power of UAVs for DS tasks $\mathbf{p}=\{p_{u,\text{LEO}}^{t,\text{sens}}|\forall u\in \mathcal{U},t\in\mathcal{T}\}$, the start time of DT tasks of UAVs $\mathbf{\delta}=\{\delta_{u}^{t,\text{tol}}|\forall u\in \mathcal{U},t\in\mathcal{T}\}$, computing resource allocation of the LEO satellite $\mathbf{f}=\{f_{u,\text{LEO}}^{t}|\forall u\in \mathcal{U},t\textcolor{blue}{\in\mathcal{T}}\}$ and offloading ratio $\mathbf{\gamma}=\{\gamma_{u,\text{LEO}}^{t}|\forall u\in \mathcal{U},t\in\mathcal{T}\}$ are jointly optimized. Mathematically, the JCORM problem can be formulated as follows:
\begin{alignat}{2}
\textbf{JCORM}\colon \mathop {{\rm{max}}}\limits_{_{\mathbf{\delta}, \mathbf{\gamma}, \mathbf{f}, \mathbf{p}}} \quad & \sum_{t=1}^{T}\Big(\sum_{u=1}^{U}{D_{u,\text{LEO}}^{t}-\omega E^{t}\Big)}\\
\mbox{s.t.}\quad
&\gamma_{u,\text{LEO}}^{t}\in[0,1], \quad \forall u,t \tag{28a}\\
&0\leq\delta_{u}^{t,\text{tol}}\leq \delta, \quad \forall u,t \tag{28b}\\
&f_{u,\text{LEO}}^{t}\in[0,F_{\text{LEO}}], \quad \forall u,t   \tag{28c}\\
&\sum_{u=1}^{U}f_{u,\text{LEO}}^{t}\leq F_{\text{LEO}}, \quad \forall t   \tag{28d}\\
&0\leq p_{u,\text{LEO}}^{t,\text{sens}}\leq p_{u,\text{LEO}}^{\text{max}}, \quad \forall u,t   \tag{28e}\\
&l_{u}^{t}\leq \delta_{u}^{t,\text{tol}}, \quad \forall u,t   \tag{28f}\\
&D_{u}^{t}\leq D_{u}^{t,\text{stor}}, \quad \forall u,t   \tag{28g}\\
&D_{u,\text{LEO}}^{t}\leq D_{u,\text{LEO}}^{t,\text{max}}, \quad \forall u,t,   \tag{28h}
\end{alignat} 
where $\omega$ is a constant that controls the relative importance of the data volume and the energy consumption. Constraint (28a) means that the offloading ratio is a binary variable. Constraint (28b) is the range limitation on the permissible start time for the DT task. Constraint (28c) is the limitation of  computing resource allocation variable. Constraint (28d) indicates that the sum of computing resources allocated by the LEO satellite to all UAVs cannot exceed the computing capacity. Constraint (28e) is the transmit power limitation of the UAV. Constraint (28f) means the latency requirement for DS tasks. It ensures that the entire processing flow for all DS tasks (including offloading, computation, and result transmission) is completed within the time slot before the UAV starts uploading DT data. This is the fundamental guarantee for the low-latency QoS requirement of DS tasks. Constraints (28g) and (28h) are storage capacity constraints essential for the DT data collection. Constraint (28g) indicates that the data volume collected by the UAV from DT-MDs in each time slot cannot exceed the remaining storage space of the UAV. Constraint (28h) ensures that in each time slot, the volume of DT data collected by the LEO satellite via each UAV does not exceed the data volume stored in the storage space of the UAV. 

\section{Algorithm Design}
In JCORM problem, constraint (28f) incorporates the maximum function. To facilitate subsequent processing, we initially decompose the maximum function in constraint (28f) into the following two equivalent constraints, 
\begin{equation}
l_{u}^{t,\text{off}}+l_{u}^{t,\text{comp}}\leq \delta_{u}^{t,\text{tol}},\quad \forall u,t,
\end{equation}
\begin{equation}
l_{u}^{t,\text{off}}+l_{u,\text{LEO}}^{t,\text{comm}}+l_{u,\text{LEO}}^{t,\text{comp}}+2l_{u,\text{LEO}}^{t,\text{prop}}\leq \delta_{u}^{t,\text{tol}},\quad \forall u,t.
\end{equation}

The problem is still a non-convex non-linear problem, which is difficult to solve directly, since it involves several continuous variables and there are coupling relations among them. Therefore, we propose a JCORM algorithm. First, the JCORM problem is decomposed into four sub-problems, i.e., transmit power optimization sub-problem, computing resource allocation sub-problem, start time optimization sub-problem of DT tasks and offloading ratio optimization sub-problem. Then, all variables are initialized. In each iteration, the four sub-problems are solved in sequence. The loop terminates when the objective value converges or the maximum number of iterations is reached.

\subsection{Transmit Power Optimization}
Given the start time of DT tasks of UAVs $\mathbf{\delta}$, computing resource allocation of the LEO satellite $\mathbf{f}$ and offloading ratio $\mathbf{\gamma}$, the terms associated with variable $\mathbf{p}$ in the objective function and constraint conditions are extracted, allowing the transmit power optimization sub-problem can be written as 
\begin{alignat}{2}
\textbf{SP1}\colon \mathop {{\rm{min}}}\limits_{_{\mathbf{p}}} \quad & \frac{\omega p_{u,\text{LEO}}^{t,\text{sens}}\gamma_{u,\text{LEO}}^{t}\sum_{k_{u}^{\text{sens}}=1}^{K_u^{\text{sens}}}D_{k_{u}^{\text{sens}}}^{t}}{R_{u,\text{LEO}}^{t,\text{sens}}}\\
\mbox{s.t.}\quad
&\text{(28e), (30)}. \notag
\end{alignat}

In SP1, the objective function is a nonlinear fraction programming problem. In order to reduce the complexity of solving the problem, the Dinkelbach method is used to convert the fraction form into the subtraction form \cite{9411713}. Let 
\begin{equation}
\eta^{r-1}=\frac{(A_u^tp_{u,\text{LEO}}^{t,\text{sens}})^{r-1}}{(R_{u,\text{LEO}}^{t,\text{sens}})^{r-1}},\quad \forall u,t,
\end{equation}
\begin{equation}
A_u^t=\omega \gamma_{u,\text{LEO}}^{t}\sum_{k_{u}^{\text{sens}}=1}^{K_u^{\text{sens}}}D_{k_{u}^{\text{sens}}}^{t}, \quad \forall u,t,
\end{equation}
where $r$ is the iteration number. Then the sub-problem SP1 can be converted to 
\begin{alignat}{2}
\textbf{SP1-1}\colon \mathop {{\rm{min}}}\limits_{_{\mathbf{p}}} \quad & A_u^tp_{u,\text{LEO}}^{t,\text{sens}}-\eta^{r-1}{R_{u,\text{LEO}}^{t,\text{sens}}}\\
\mbox{s.t.}\quad
&\text{(28e)}, \text{(30)}. \notag
\end{alignat}

The optimization problem SP1-1 mentioned above is convex with respect to (w.r.t.) the variable $\mathbf{p}$ under the given constraints. Consequently, we employ the Lagrange multiplier method to solve it. The Lagrangian function \cite{9126800} for the above problem can be expressed as follows: 
\begin{equation}
\begin{aligned}
&\quad L(p_{u,\text{LEO}}^{t,\text{sens}},\lambda_{u,\text{LEO}}^{t,\text{sens}},\mu_{u,\text{LEO}}^{t,\text{sens}})\\
&=A_u^tp_{u,\text{LEO}}^{t,\text{sens}}-\eta^{r-1}R_{u,\text{LEO}}^{t,\text{sens}}\\&\quad+\lambda_{u,\text{LEO}}^{t,\text{sens}}(\gamma_{u,\text{LEO}}^{t}\sum_{k_{u}^{\text{sens}}=1}^{K_u^{\text{sens}}}D_{k_{u}^{\text{sens}}}^{t}-\Gamma_u^{t}R_{u,\text{LEO}}^{t,\text{sens}})\\
&\quad+\mu_{u,\text{LEO}}^{t,\text{sens}}(p_{u,\text{LEO}}^{t,\text{sens}}-p_{u,\text{LEO}}^{\text{max}}),
\quad \forall u,t,
\end{aligned}
\end{equation}
where 
\begin{equation}
\Gamma_u^{t}=\delta_{u}^{t,\text{tol}}-2l_{u,\text{LEO}}^{t,\text{prop}}-l_{u,\text{LEO}}^{t,\text{comp}}-l_{u}^{t,\text{off}},\quad \forall u,t.
\end{equation}

By differentiating the Lagrangian function w.r.t. the variable $\mathbf{p}$, we obtain
\begin{equation}
\begin{aligned}
&\quad \frac{\partial L}{\partial p_{u,\text{LEO}}^{t,\text{sens}}}\!=\!\omega \gamma_{u,\text{LEO}}^{t}\sum_{k_{u}^{\text{sens}}=1}^{K_u^{\text{sens}}}D_{k_{u}^{\text{sens}}}^{t}\!-\!\frac{\eta^{r-1}B_{\text{LEO}}g_{u,\text{LEO}}^{t}}{U\ln2(\sigma^2+p_{u,\text{LEO}}^{t,\text{sens}}g_{u,\text{LEO}}^{t})}\\&\quad\quad\quad\quad\quad\!-\!\frac{\lambda_{u,\text{LEO}}^{t,\text{sens}}\Gamma_u^{t}B_\text{LEO}g_{u,\text{LEO}}^{t}}{U\ln2(\sigma^2+p_{u,\text{LEO}}^{t,\text{sens}}g_{u,\text{LEO}}^{t})}+\mu_{u,\text{LEO}}^{t,\text{sens}},
\quad \forall u,t.
\end{aligned}
\end{equation}

Let $\partial L(p_{u,\text{LEO}}^{t,\text{sens}},\lambda_{u,\text{LEO}}^{t,\text{sens}},\mu_{u,\text{LEO}}^{t,\text{sens}})/{\partial p_{u,\text{LEO}}^{t,\text{sens}}}=0$, then the optimal transmit power of the UAV for DS tasks is 
\begin{equation}
\begin{aligned}
&\quad \hat{p}_{u,\text{LEO}}^{t,\text{sens}}\\&=\Big\{\frac{\eta^{r-1}B_{\text{LEO}}+\lambda_{u,\text{LEO}}^{t,\text{sens}}\Gamma_u^{t}B_\text{LEO}}{{(\omega \gamma_{u,\text{LEO}}^{t}\sum_{k_{u}^{\text{sens}}=1}^{K_u^{\text{sens}}}D_{k_{u}^{\text{sens}}}^{t}+\mu_{u,\text{LEO}}^{t,\text{sens}})U\ln2}}\\&\qquad-\frac{\sigma^2}{g_{u,\text{LEO}}^{t}}\Big\}^{+},\quad \forall u,t,
\end{aligned}
\end{equation}
where the function $\{\mathbf{x}\}^+=\max\{\mathbf{x},0\}$.

The Lagrange multipliers $\lambda_{u,\text{LEO}}^{t,\text{sens}}$ and $\mu_{u,\text{LEO}}^{t,\text{sens}}$ can be solved by using the sub-gradient method. In the $(j+1)$-th iteration of the sub-gradient method, $\lambda_{u,\text{LEO}}^{t,\text{sens}}(j+1)$ and $\mu_{u,\text{LEO}}^{t,\text{sens}}(j+1)$ can be updated according to the following equations:
\begin{equation}
\lambda_{u,\text{LEO}}^{t,\text{sens}}(j+1)=\{\lambda_{u,\text{LEO}}^{t,\text{sens}}(j)-\alpha^{\lambda}(j)\nabla_{\lambda}L(j)\}^{+},\quad \forall u,t,j,
\end{equation}
\begin{equation}
\mu_{u,\text{LEO}}^{t,\text{sens}}(j+1)=\{\mu_{u,\text{LEO}}^{t,\text{sens}}(j)-\alpha^{\mu}(j)\nabla_{\mu}L(j)\}^{+},\quad \forall u,t,j,
\end{equation}
here, $\alpha^{\lambda}(j)$ and $\alpha^{\mu}(j)$ represent the step sizes at the $j$-th iteration, and the corresponding sub-gradients can be obtained as follows:
\begin{equation}
\nabla_{\lambda}L(j)=\gamma_{u,\text{LEO}}^{t}\sum_{k_{u}^{\text{sens}}=1}^{K_u^{\text{sens}}}D_{k_{u}^{\text{sens}}}^{t}-\Gamma_u^{t}{\hat{R}}_{u,\text{LEO}}^{t,\text{sens}},\quad \forall u,t,j,
\end{equation}
\begin{equation}
\nabla_{\mu}L(j)=\hat{p}_{u,\text{LEO}}^{t,\text{sens}}(j)-p_{u,\text{LEO}}^{\text{max}}, \quad \forall u,t,j,
\end{equation}
where $\hat{p}_{u,\text{LEO}}^{t,\text{sens}}(j)$ represents the optimal transmit power at the $j$-th iteration, which is derived from the Equation (38). The proposed algorithm based on the Dinkelbach method is summarized as in \textbf{Algorithm \ref{alg:Algorithm 1}}.

    \begin{algorithm}[t]
      \caption{Dinkelbach Method-based Power Optimization Algorithm}
        \label{alg:Algorithm 1}
          \LinesNumbered
        {
        \textbf{Initialization}: $(p_{u,\text{LEO}}^{t,\text{sens}})^r$, the maximum iteration number $r_{\max}$, $r=1$, the tolerance $\epsilon$, and $\eta^0$. \\
        \textbf{While} $\lvert(A_u^tp_{u,\text{LEO}}^{t,\text{sens}}-\eta^{r}{R_{u,\text{LEO}}^{t,\text{sens}}})-(A_u^tp_{u,\text{LEO}}^{t,\text{sens}}-\eta^{r-1}{R_{u,\text{LEO}}^{t,\text{sens}}})\rvert>\epsilon$ or $r=r_{\max}$ \textbf{do}\\
        \quad Given the threshold $\xi$, inner-layer maximum iteration number $j_{\max}$, inner-layer iteration number $j=1$, Lagrange multipliers $\lambda_{u,\text{LEO}}^{t,\text{sens}}(j)$ and $\mu_{u,\text{LEO}}^{t,\text{sens}}(j)$;\\
     \quad   \textbf{Repeat}\\
          \quad\quad Calculate optimal power $(p_{u,\text{LEO}}^{t,\text{sens}}(j))^r$ according to (38);\\
          \quad\quad Update $j=j+1$;\\
          \quad\quad Update Lagrange multipliers $(\lambda_{u,\text{LEO}}^{t,\text{sens}}(j))^r$ and $(\mu_{u,\text{LEO}}^{t,\text{sens}}(j))^r$ according to (39) and (40);\\
          }
      \quad   \textbf{Until} Convergence or the iteration time $j=j_{\max}$\\
     \quad Update $(p_{u,\text{LEO}}^{t,\text{sens}})^r=(p_{u,\text{LEO}}^{t,\text{sens}}(j))^r$;\\
      \quad Update $r=r+1$.\\
       \textbf{End}\\
       \textbf{Return} The transmit power of UAVs for DS tasks.
    \end{algorithm}

\subsection{Computing Resource Allocation}
Given the start time of DT tasks of UAVs $\mathbf{\delta}$, the transmit power $\mathbf{p}$ and offloading ratio $\mathbf{\gamma}$, the terms associated with variable $\mathbf{f}$ in the objective function and constraints of original problem (28) are extracted, allowing the computing resource allocation sub-problem can be written as 
\begin{alignat}{2}
\textbf{SP2}\colon \mathop {{\rm{max}}}\limits_{_{\mathbf{f}}} \quad & \sum_{t=1}^{T}\Big(\sum_{u=1}^{U}{D_{u,\text{LEO}}^{t}-\omega E^{t}\Big)}\\
\mbox{s.t.}\quad
&(28\text{c}),(28\text{d}),(30). \notag
\end{alignat}

It can be observed that \textbf{SP2} is a quadratic convex problem w.r.t. variable $\mathbf{f}$. The objective function of \textbf{SP2} is a monotonic increasing function in the domain. The optimal computing resource allocation variable can be obtained at its lower bound \cite{9515574}. Its closed-form solution is given by
\begin{equation}
\hat{f}_{u,\text{LEO}}^{t}=f_{u,\text{LEO}}^{t,\min}=\frac{f_0 \gamma_{u,\text{LEO}}^{t}\sum_{k_{u}^{\text{sens}}=1}^{K_u^{\text{sens}}}D_{k_{u}^{\text{sens}}}^{t}}{\Gamma_u^{t}},\quad \forall u,t.
\end{equation}

\subsection{The Start Time Optimization of DT tasks}
Given the computing resource allocation variable $\mathbf{f}$, the transmit power $\mathbf{p}$ and offloading ratio $\mathbf{\gamma}$, the terms associated with variable $\mathbf{\delta}$ in the objective function and constraints of original problem (28) are extracted, allowing the start time optimization sub-problem can be written as 
\begin{alignat}{2}
\textbf{SP3}\colon \mathop {{\rm{max}}}\limits_{_{\mathbf{\delta}}} \quad & \sum_{t=1}^{T}\Big(\sum_{u=1}^{U}{D_{u,\text{LEO}}^{t}-\omega E^{t}\Big)}\\
\mbox{s.t.}\quad
&(28\text{b}),(28\text{g}),(28\text{h}),(29),(30). \notag
\end{alignat}

By combining constraints (29) with constraints (30), and combining constraints (28g) with constraints (28h), the optimization problem above is rewritten as
\begin{alignat}{2}
\textbf{SP3-1}\colon \mathop {{\rm{max}}}\limits_{_{\mathbf{\delta}}} \quad & \sum_{t=1}^{T}\Big(\sum_{u=1}^{U}{D_{u,\text{LEO}}^{t}-\omega E^{t}\Big)}\\
\mbox{s.t.}\quad
&(28\text{b}) \notag\\
&l_u^{t,\text{off}}+l_{u,\text{LEO}}^{t,\text{prop}}+\Lambda_u^{t,1} \leq \delta_u^{t,\text{tol}},\quad\forall u,t\tag{46a}\\
&\Lambda_u^{t,2}\leq  \delta_u^{t,\text{tol}}\leq \frac{D_u^{t,\text{stor}}}{\sum_{k_{u}^{\text{tol}}=1}^{K_u^{\text{tol}}}R_{k_{u}^{\text{tol}},u}^{t}}, \forall u,t,\tag{46b}
\end{alignat}
where
\begin{equation}
\begin{aligned}
\quad\Lambda_u^{t,1}&=\frac{1}{2}\sum_{k_{u}^{\text{sens}}=1}^{K_u^{\text{sens}}}D_{k_{u}^{\text{sens}}}^{t}\\&\quad\times(\frac{f_0}{F_u}+\gamma_{u,\text{LEO}}^{t}\big(\frac{1}{R_{u,\text{LEO}}^{t,\text{sens}}}+\frac{f_0}{f_{u,\text{LEO}}^{t}}-\frac{f_0}{F_u}\big)),\quad\forall u,t,
\end{aligned}
\end{equation}
\begin{equation}
\Lambda_u^{t,2}=\frac{R_{u,\text{LEO}}^{t,\text{tol}}\delta-(D_{u}^{\text{stor}}-D_{u}^{t,\text{stor}})}{R_{u,\text{LEO}}^{t,\text{tol}}+\sum_{k_{u}^{\text{tol}}=1}^{K_u^{\text{tol}}}R_{k_{u}^{\text{tol}},u}^{t}},  \quad \forall u,t.
\end{equation}

    \begin{algorithm}[t]
      \caption{JCORM Algorithm}
        \label{alg:Algorithm 2}
          \LinesNumbered
        {
        \textbf{Initialization}: $\mathbf{\delta}$, $\mathbf{\gamma}$, $\mathbf{p}$, $\mathbf{f}$ , the maximum iteration number $i_{\max}$, the iteration time $i=1$, the tolerance $\tau$, and the number of time slots $T$. \\
        \textbf{For} $t=1:T$\\
     \quad   \textbf{Repeat}\\
          \qquad Obtain $\hat{p}_{u,\text{LEO}}^{t,\text{sens}}$ by using the \textbf{Algorithm \ref{alg:Algorithm 1}};\\
          \qquad Obtain $\hat{f}_{u,\text{LEO}}^{t}$ by calculating (44);\\
         \qquad Obtain $\hat{\delta}_{u}^{t,\text{tol}}$ by solving the problem \textbf{SP3-1};\\
         \qquad Obtain $\hat{\gamma}_{u,\text{LEO}}^{t}$ by calculating (54);\\
         \qquad Update $i=i+1$;\\
          }
      \quad   \textbf{Until} Convergence or the iteration time $i=i_{\max}$\\
     \quad Update the remaining storage space of UAVs according to (22);\\
      \quad Update $t=t+1$;\\
       \textbf{End}\\
       \textbf{Return} The utility of CSAMN.
    \end{algorithm}

Obviously, the optimization problem \textbf{SP3-1} is a linear optimization problem. By analyzing the form of its function curve, there are two cases:

\subsubsection{Case I} If $\omega p_{u,\text{LEO}}^{t,\text{tol}}-R_{u,\text{LEO}}^{t,\text{tol}}\geq0$, then the objective function in problem SP3-1 is monotonically increasing. Therefore, when $\delta_u^{t,\text{tol}}$ takes the maximum value, the objective value is the largest, i.e.,
\begin{equation}
\hat{\delta}_u^{t,\text{tol}}=\min\Big\{\delta, \frac{D_u^{t,\text{stor}}}{\sum_{k_{u}^{\text{tol}}=1}^{K_u^{\text{tol}}}R_{k_{u}^{\text{tol}},u}^{t}}\Big\},  \quad\forall u,t.
\end{equation}

\subsubsection{Case II} If $\omega p_{u,\text{LEO}}^{t,\text{tol}}-R_{u,\text{LEO}}^{t,\text{tol}}<0$, then the objective function in problem \textbf{SP3-1} is monotonically decreasing. Therefore, when $\delta_u^{t,\text{tol}}$ takes the minimum value, the objective value is the largest, i.e.,
\begin{equation}
\hat{\delta}_u^{t,\text{tol}}=\max\{0,\Lambda_u^{t,2},l_u^{t,\text{off}}+l_{u,\text{LEO}}^{t,\text{prop}}+\Lambda_u^{t,1}\},  \quad\forall u,t.
\end{equation}

\subsection{Offloading Ratio Optimization}
Given the start time of DT tasks of UAVs $\mathbf{\delta}$, the computing resource allocation variable $\mathbf{f}$ and the transmit power $\mathbf{p}$, the terms associated with variable $\mathbf{\gamma}$ in the objective function and constraints of original problem (28) are extracted, allowing the offloading ratio optimization sub-problem can be written as 
\begin{alignat}{2}
\textbf{SP4}\colon \mathop {{\rm{max}}}\limits_{_{\mathbf{\gamma}}} \quad & \sum_{t=1}^{T}\Big(\sum_{u=1}^{U}{D_{u,\text{LEO}}^{t}-\omega E^{t}\Big)}\\
\mbox{s.t.}\quad
&(28\text{a}),(29),(30). \notag
\end{alignat}

Similarly, this is a linear optimization problem. Firstly, we rewrite the constraint in the following form by conbining all constraints,
\begin{equation}
\begin{aligned}
&\gamma_{u,\text{LEO}}^{t,\min}=\max\Big\{0,\frac{l_u^{t,\text{off}}F_u-\delta_u^{t,\text{tol}}F_u}{f_0\sum_{k_{u}^{\text{sens}}=1}^{K_u^{\text{sens}}}D_{k_{u}^{\text{sens}}}^{t}}+1\Big\}  \leq \gamma_{u,\text{LEO}}^{t}\\&\leq \min\Big\{\frac{\delta_u^{t,\text{tol}}-2l_{u,\text{LEO}}^{t,\text{prop}}-l_{u}^{t,\text{off}}}{\big(\frac{1}{R_{u,\text{LEO}}^{t,\text{sens}}}+\frac{f_0}{f_{u,\text{LEO}}^{t}}\big)\sum_{k_{u}^{\text{sens}}=1}^{K_u^{\text{sens}}}D_{k_{u}^{\text{sens}}}^{t}},1\Big\}= \gamma_{u,\text{LEO}}^{t,\max}, \quad\forall u,t.
\end{aligned}
\end{equation}

Moreover, the linear coefficient of the objective function is
\begin{equation}
\begin{aligned}
\Pi_u^t&=\omega \sum_{k_{u}^{\text{sens}}=1}^{K_u^{\text{sens}}}D_{k_{u}^{\text{sens}}}^{t} \\&\quad\times\Big(f_0\kappa F_u^2-f_0\kappa (f_{u,\text{LEO}}^t)^2-\frac{p_{u,\text{LEO}}^{t,\text{sens}}}{R_{u,\text{LEO}}^{t,\text{sens}}}\Big), \quad\forall u,t.
\end{aligned}
\end{equation}

Therefore, the offloading ratio can be obtained according to the following formula
\begin{equation}
\hat{\gamma}_{u,\text{LEO}}^{t}=\left\{
\begin{array}{rcl}
\gamma_{u,\text{LEO}}^{t,\min} &    &\Pi_{u}^{t}\leq 0,\\
\gamma_{u,\text{LEO}}^{t,\max} &    &\Pi_{u}^{t}> 0.
\end{array}
\right.
\end{equation}

\subsection{Joint Computation Offloading and Resource Management Algorithm}
Based on the above analyses, we propose a JCORM algorithm to solve the optimization problem (28), where $\mathbf{\delta}$, $\mathbf{\gamma}$, $\mathbf{p}$ and $\mathbf{f}$ are alternatively optimized. The detailed steps of the proposed algorithm are listed in \textbf{Algorithm \ref{alg:Algorithm 2}}. 

\textit{Convergence Analysis}: In each iteration of JCORM Algorithm, we sequentially optimize four variables. Specifically, SP1 is transformed into a convex problem by the Dinkelbach method and is solved optimally using the Lagrange multiplier method. Then, SP2, SP3, and SP4 are solved by deriving their closed-form solutions. These solutions are optimal for each sub-problem when other variables are fixed. Since each step of the alternating optimization independently improves the objective function (28) or keeps it unchanged, the series of objective values generated by the JCORM algorithm are non-decreasing as the number of iterations increases. Moreover, the objective function is bounded within a finite time slot. According to the monotone convergence theorem, the proposed JCORM algorithm guarantees convergence to a local optimal solution.

\textit{Complexity Analysis}: In \textbf{Algorithm \ref{alg:Algorithm 1}}, it needs $U$ times calculations for calculation of transmit power, thus the complexity can be denoted by $\mathcal{O}(\hat{r}\hat{j}U)$, where $\hat{r}$ and $\hat{j}$ denote the number of iterations required for the inner and outer loops of \textbf{Algorithm \ref{alg:Algorithm 1}} to reach convergence, respectively. In addition, updates to variables start time for DT tasks, computing resource allocation, and offloading ratio all require $U$ calculations. Let $\hat{i}$ denotes the number of iterations required for JCORM Algorithm, then the complexity is given by $\mathcal{O}((\hat{r}\hat{j}U+3U)T\hat{i})$.

\section{Performance Evaluation}
We conduct a couple of simulations to validate the proposed algorithms and analyze the performance. 
\subsection{Simulation Setup}
In this section, we list some simulation experiments to present the performance of the proposed solution. $U=6$ UAVs are uniformly distributed within the network range of 2 km$\times$2 km and the altitude of the UAV is 500 m. The altitude of the LEO satellite is 780 km. For each UAV, the number of DS-MDs and DT-MDs are randomly distributed between [1, 5] and [5, 10], respectively. The length of the time slot is $\delta=10$ s. The bandwidth of the UAV and the LEO satellite are 10 MHz and 40 MHz, respectively. The maximum transmit power of the UAV is 1 W. The data volume of DS tasks of each DS-MD is randomly distributed between [1, 3] Mbits. The CPU cycles required to process 1 bit data is 400 cycles. The computing capabilities of the UAV and the LEO satellite are 2 GHz and 10 GHz, respectively. The Rician factor is 10. Other simulation parameters are listed in TABLE ~\ref{table 1}. Some parameters are derived from the references \cite{9184929} \cite{Jung_2023} \cite{10764739} \cite{9515574}.

In order to evaluate the performance of the proposed algorithm, we compare it with three benchmarks: 

\textit{Average time slot method (ATSM):} A fixed time allocation strategy is adopted, where the time of each time slot is evenly distributed between the DS tasks and the DT tasks. Other variables are obtained through algorithm optimization.

\textit{Successive convex approximation (SCA)-based algorithm:} The SCA method is employed to solve the transmit power optimization sub-problem. Specifically, the original transmit power optimization sub-problem is iteratively approximated as a series of convex problems. These convex problems are then efficiently solved using the CVX solver. Other variables are obtained through algorithm optimization.

\textit{Heuristic algorithm (HA):} The JCORM problem is directly solved using the genetic algorithm. As a heuristic optimization method, it is employed to compare the efficiency of the alternating optimization algorithm proposed by this paper.

In this section, we evaluate the performance of the proposed JCORM algorithm through extensive simulations. The following key performance metrics are used for assessment:

\textit{Utility:} The primary objective of our optimization, defined in Eq. (28), which captures the trade-off between the data collection capability and energy consumption of CSAMN.

\textit{Total data volume:} The cumulative amount of DT data, successfully collected by the LEO satellite, reflecting the benefit gained by CSAMN system.

\textit{Total energy consumption:}The overall energy cost, consumed by the UAVs and the LEO satellite for communication and computation, as detailed in Eqs. (23)-(27).

\textit{Average delay for DS tasks:} The average delay is defined as the average processing time of all UAVs handling DS tasks across all time slots. This metric is crucial for verifying that the latency requirements of DS tasks are met.

We compare the proposed JCORM algorithm with three benchmarks and analyze the impact of key network parameters on the above metrics.
\begin{table}
\renewcommand{\arraystretch}{1.3}
\caption{Simulation Parameters}
\label{table 1}
\centering
\begin{tabular}{|c|c|c|c|}
\hline
\textbf{Parameter} & \textbf{Value} & \textbf{Parameter} & \textbf{Value}\\
\hline
$g_0$ & $-30$ dB & $c$ & $3\times10^8$ \text{m/s}\\
\hline
$\sigma^2$ & $-80$ dBm & $\kappa$ & $10^{-28}$\\
\hline
$G$ & $10$ dB & $\omega$  & $10$\\
\hline
$p_{k_{u}^\text{tol},u}^{t}$ & $0.3$ W & $\beta_u^t$ & $0.6$\\
\hline
$p_{k_{u}^\text{sens},u}^{t}$ & $0.3$ W & $r_{\text{E}}$ & $6371$ km\\
\hline
$D_{u}^{0,\text{stor}}$ & $1$ GB & $\theta$ & $20^\circ$\\
\hline
$D_{u}^{\text{stor}}$ & $1.5$ GB & $v_s$  & $7.5$ km/s\\
\hline
$PL_{\text{c}}$& $1$& $PL_{\text{e}}$& $2$ \\
\hline
$r_{\max}$, $j_{\max}$, $i_{\max}$& $50$ & $\epsilon$, $\xi$, $\tau$ & $0.01$\\
\hline
\end{tabular}
\end{table}
\vspace{6pt}
\subsection{Performance Analysis}
\begin{figure*}
	\setlength{\abovecaptionskip}{-5pt}
	\setlength{\belowcaptionskip}{-10pt}
	\centering
	\begin{minipage}[t]{0.33\linewidth}
		\centering
		\includegraphics[width=2in]{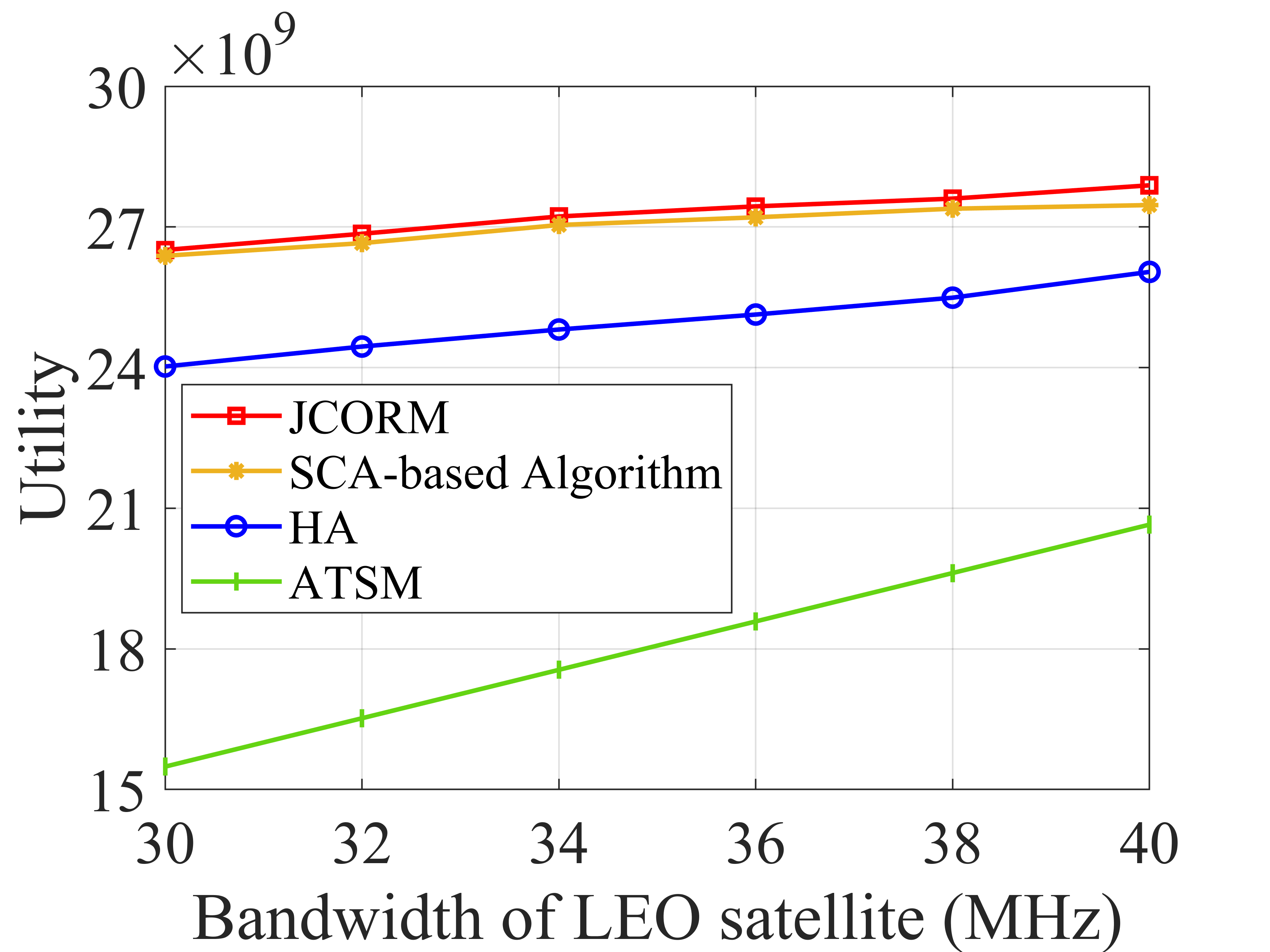}
        \centerline{(a)}
	\end{minipage}%
	\begin{minipage}[t]{0.33\linewidth}
		\centering
		\includegraphics[width=2in]{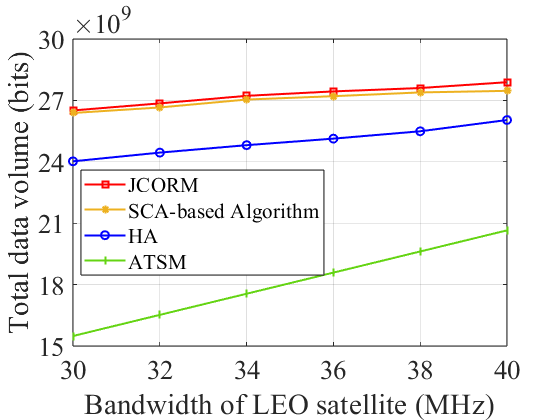}
 \centerline{(b)}
	\end{minipage}
	\begin{minipage}[t]{0.33\linewidth}
		\centering
		\includegraphics[width=2in]{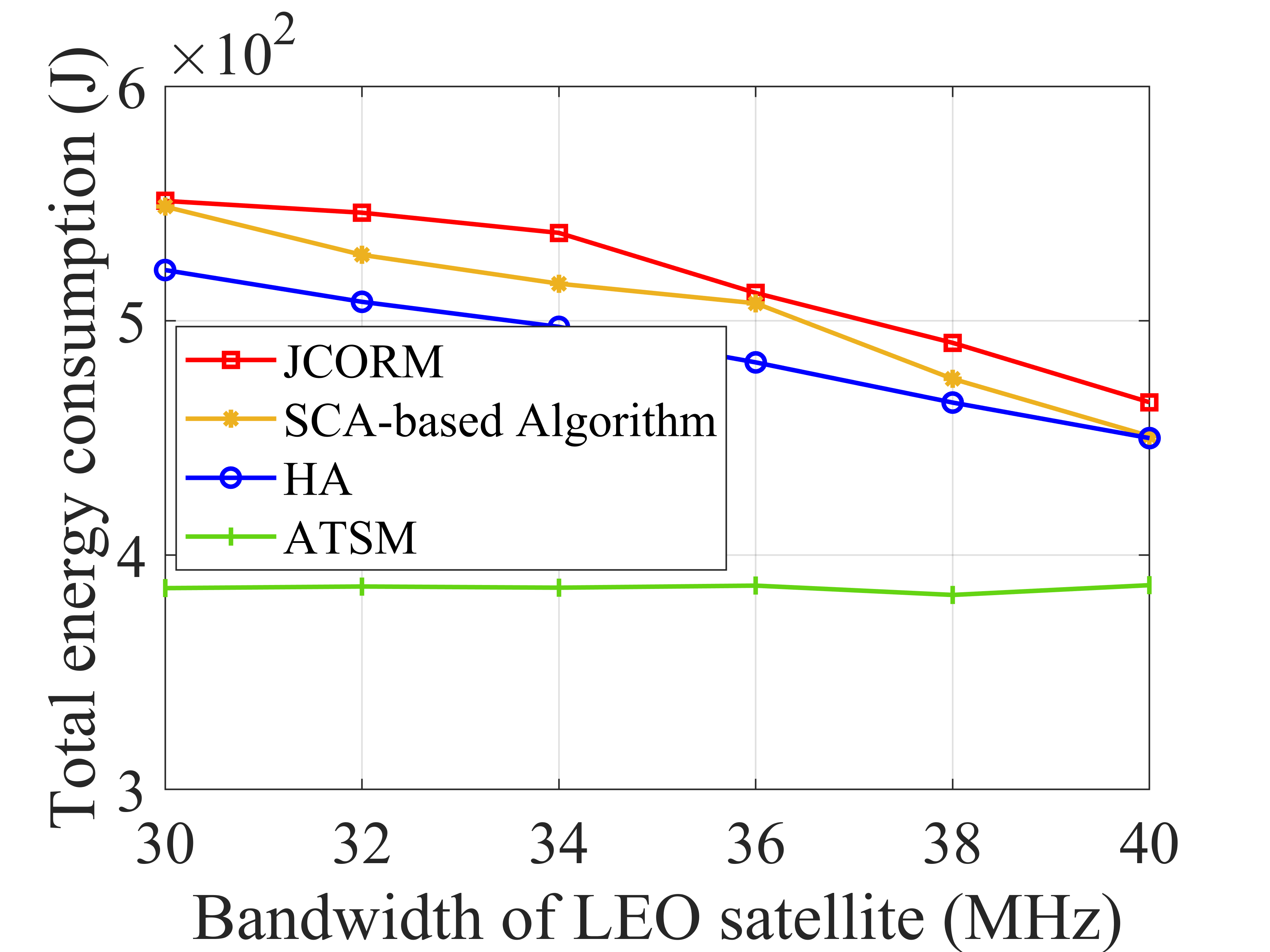}
              \centerline{(c)}
	\end{minipage}
    \vspace{5pt} 
    
        		\caption{Effect of the bandwidth of LEO satellite: (a) utility, (b) total data volume, and (c) total energy consumption.}
		\label{Fig.3}
\end{figure*}

Fig. \ref{Fig.3} illustrates the impact of LEO satellite bandwidth on system performance, including utility, total data volume, and total energy consumption. As observed in Fig. \ref{Fig.3}(a), the utility increases with the expansion of the LEO satellite bandwidth. This improvement arises from the increased transmission rates for both DS data and DT data from the UAV to the LEO satellite, facilitated by the expanded bandwidth. Specifically, for DS data transmission, the increased transmission rate significantly reduces the offloading delay, thereby decreasing the processing time of DS data within each time slot. This reduction allows for more time allocated to DT data transmission, which consequently elevates utility. Higher transmission rates enable LEO satellites to collect increased DT data volumes per time slot, directly enhancing utility. Fig. \ref{Fig.3}(b) further confirms that the total data volume collected by the LEO satellite grows significantly as the bandwidth increases, underscoring the positive correlation between bandwidth and data throughput. In particular, as demonstrated in Fig. \ref{Fig.3}(c), the total energy consumption decreases with increasing bandwidth. The reason is that the higher data rate significantly reduces the time required for transmitting both DS tasks and DT data. This reduction in energy consumption, coupled with the surge in collected data volume, synergistically contributes to the remarkable improvement in the utility.

\begin{figure}[t]
\centering 
{\includegraphics[width=0.35\textwidth]{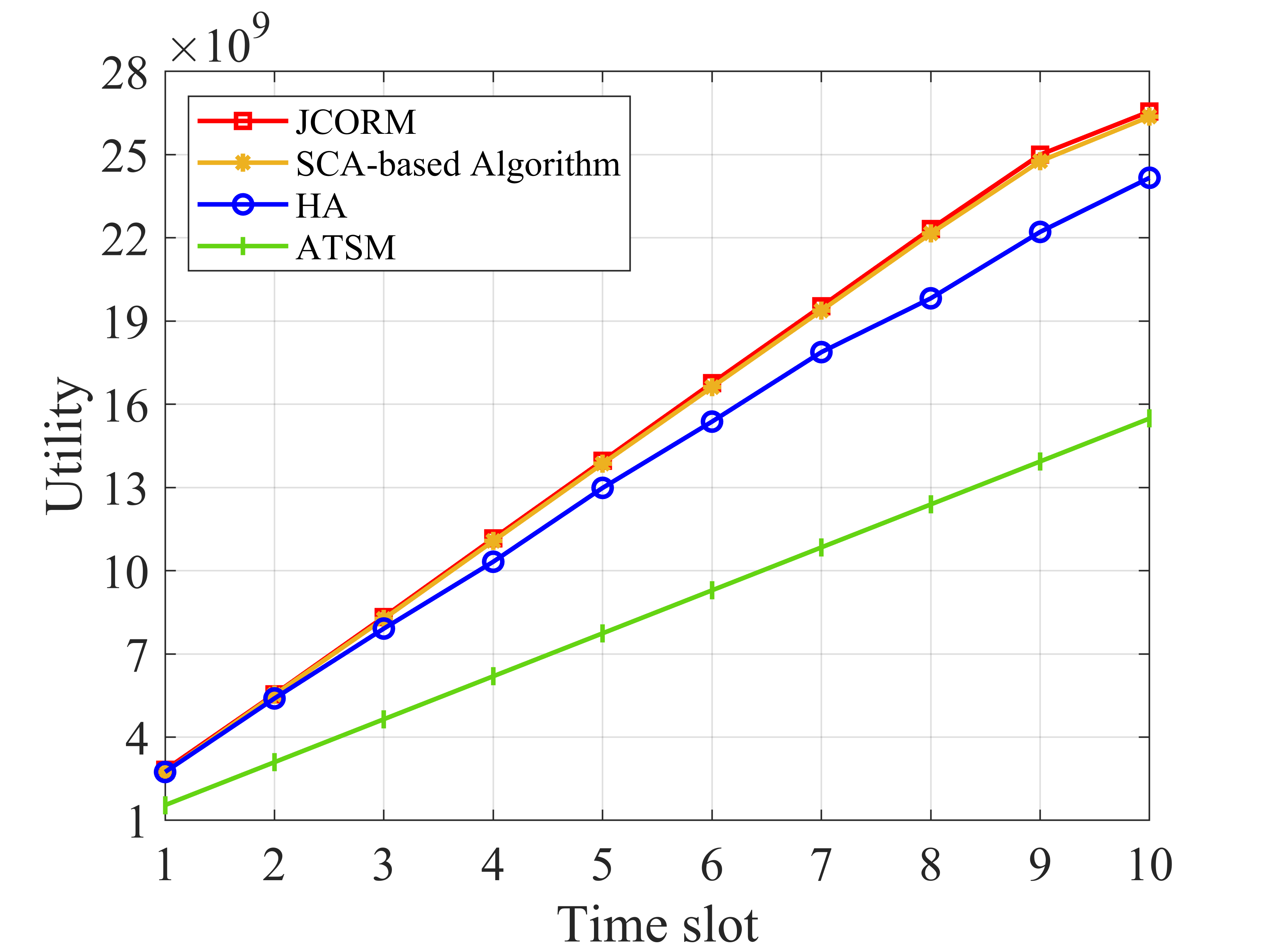}}
\vspace{-3pt} 
\caption{Utility with the index of time slots.}
\label{Fig.4}
\end{figure}
\begin{table}
\renewcommand{\arraystretch}{1.2}
\caption{Complexity Comparison between JCORM Algorithm and SCA-based Algorithm}
\label{table 2}
\centering
\begin{tabular}{|c|c|c|}
\hline
\textbf{Algorithm} & \textbf{Complexity} & \textbf{\makecell[c]{Running time for an  \\ experiment (second)}}\\
\hline
JCORM Algorithm & $\mathcal{O}((\hat{r}\hat{j}U+3U)T\hat{i})$ & $0.16$ \\
\hline
SCA-based Algorithm & $\mathcal{O}((L_sU^{3}+3U)T\hat{i})$ & $318.21$ \\
\hline
\end{tabular}
      \begin{tablenotes}  
        \item *$L_s$ is the number of iteration required of SCA-based Algorithm to reach convergence. 
       \end{tablenotes}  
\end{table}

Fig. \ref{Fig.4} illustrates the relationship between the utility and the index of time slots. It can be seen that as the index of time slots increases, the utility of the network grows almost linearly. Evidently, the proposed JCORM algorithm outperforms the other three methods. After completing data collection over ten time slots, the utility of the JCORM algorithm is approximately increased by 0.7\%, 8.9\%, and 41.5\% compared to SCA-based Algorithm, HA, and ATSM, respectively. Unlike the joint optimization capability of the JCORM scheme, ATSM adopts a fixed time slot allocation strategy, which limits the time for LEO satellite to collect DT data, thereby significantly reducing the utility. Moreover, the utility obtained by the SCA-based Algorithm is very close to that of the JCORM algorithm. This is because the SCA-based Algorithm transforms the power optimization sub-problem into a series of convex problems, enabling it to effectively solve for an approximately optimal transmit power. This also illustrates the effectiveness of the JCORM algorithm. However, as shown in TABLE \ref{table 2}, JCORM significantly outperforms the SCA-based Algorithm in terms of running time and computational complexity.
\begin{figure}
	\setlength{\abovecaptionskip}{-5pt}
	\setlength{\belowcaptionskip}{-10pt}
	\centering
	\begin{minipage}[t]{0.5\linewidth}
		\centering
		\includegraphics[width=1.1\linewidth]{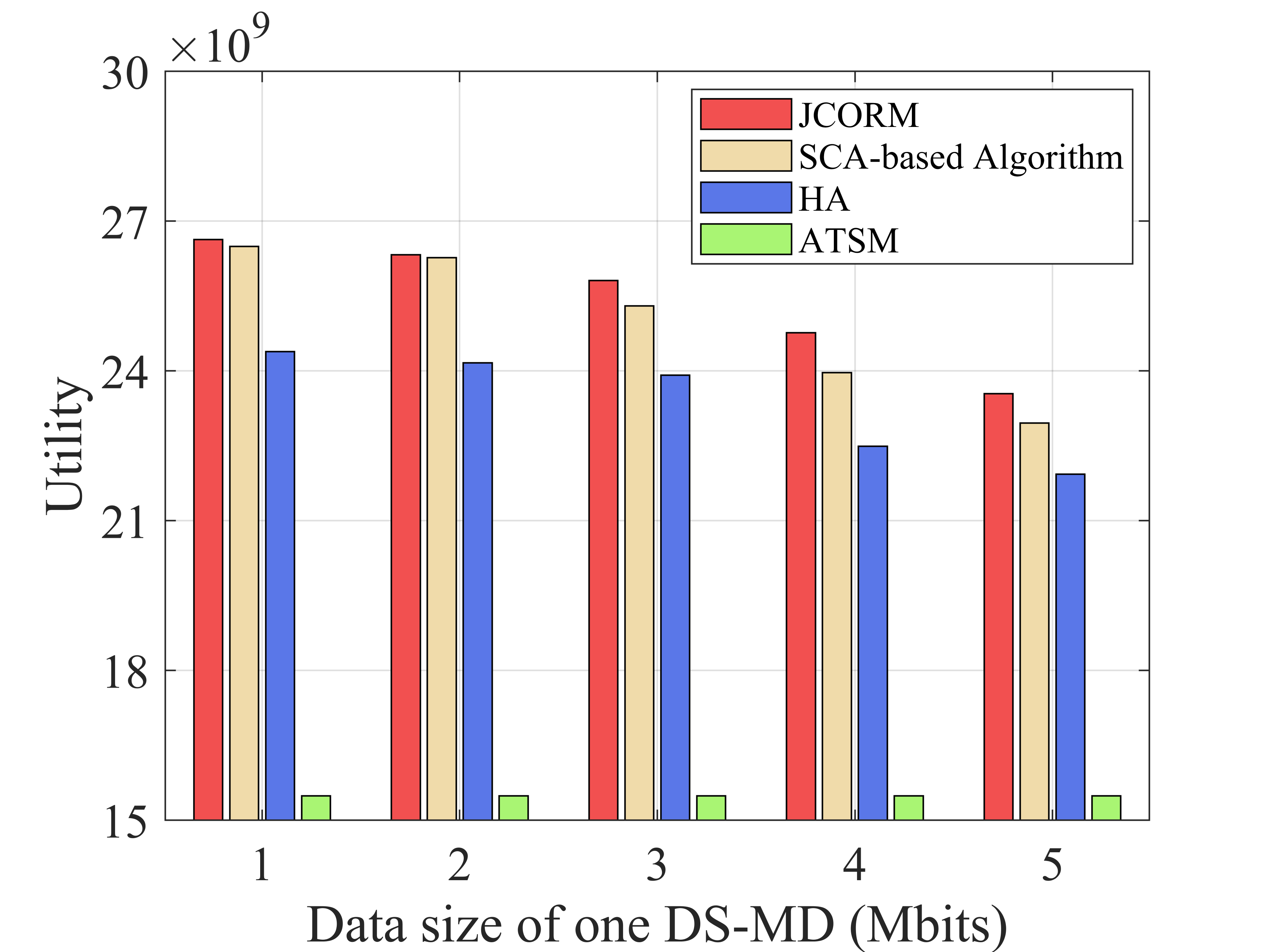}
        \centerline{(a)}
	\end{minipage}%
	\begin{minipage}[t]{0.5\linewidth}
		\centering
		\includegraphics[width=1.1\linewidth]{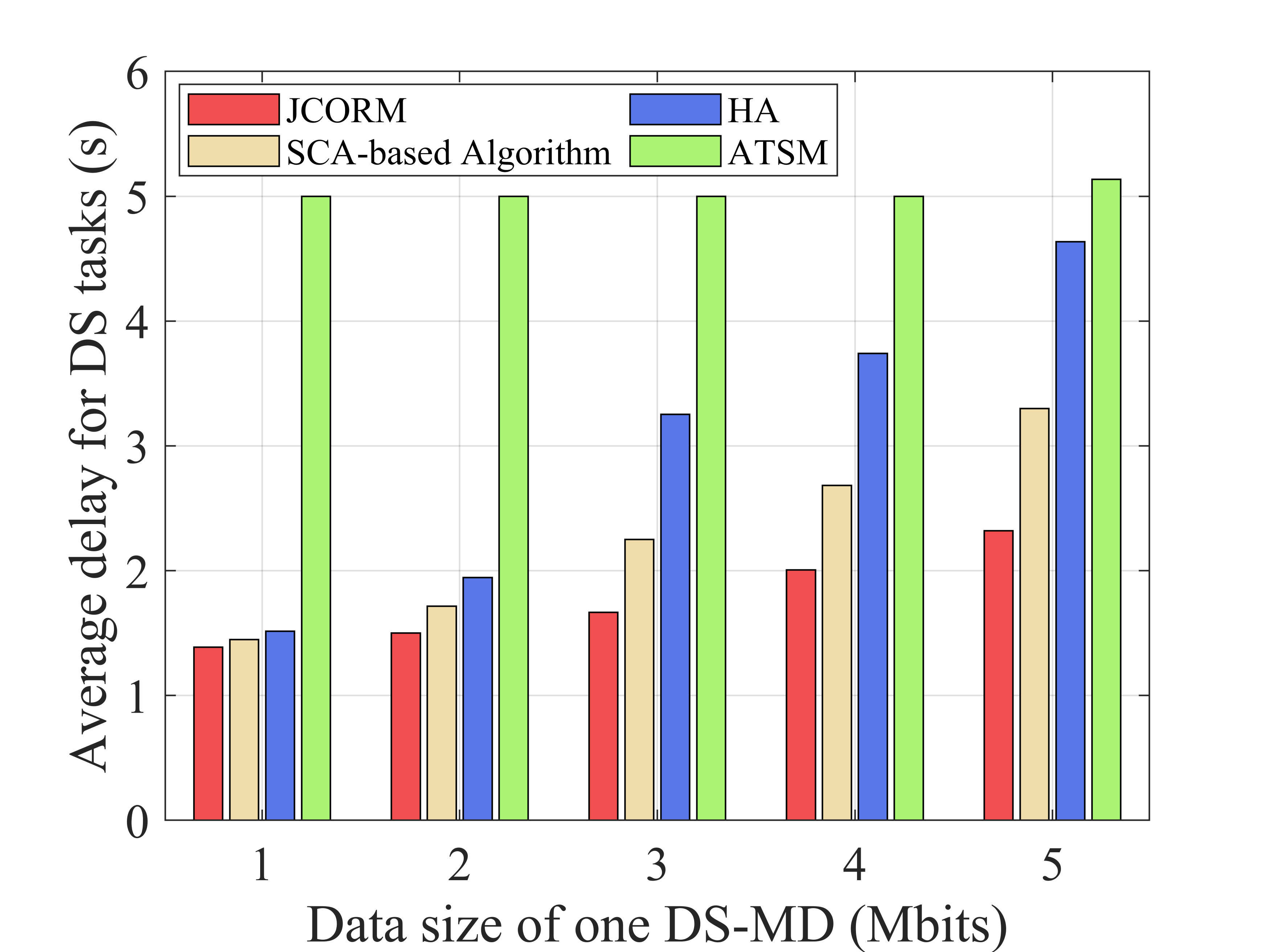}
 \centerline{(b)}
	\end{minipage}
    \vspace{6pt} 
      \caption{The relationship between network performance and data size of one DS-MD: (a) utility vs. data size of one DS-MD and (b) average delay of DS tasks vs. data size of one DS-MD.}
		\label{Fig.5}
\end{figure}

In  Fig. \ref{Fig.5}(a), it is shown that for the proposed JCORM algorithm, SCA-based Algorithm, and HA, an increase in the data size of DS-MDs leads to a decrease in utility. Since the larger data size of DS-MDs results in a longer processing time for DS tasks in each time slot, thereby reducing the time available for DT data collection and ultimately leading to a decrease in the utility. The processing time allocated to DS tasks and DT tasks in each time slot remains unchanged in ATSM. Therefore, utility of ATSM is not affected by the data volume of DS-MDs. Correspondingly, Fig. \ref{Fig.5}(b) analyzes the variation in the average delay of DS tasks under different data size of DS-MDs. It can be seen that the proposed JCORM algorithm, through joint optimization strategy, can significantly reduce the processing delay of DS tasks. For ATSM, when the data size of one DS-MD is 5 Mbits, the average delay of DS tasks is higher than setting a half-time slot. In this case, compared with the SCA-based Algorithm, HA and ATSM, the average delay of JCORM is reduced by 29.6\%, 49.9\%, and 54.9\%, respectively.
\begin{figure}
	\setlength{\abovecaptionskip}{-5pt}
	\setlength{\belowcaptionskip}{-10pt}
	\centering
	\begin{minipage}[t]{0.5\linewidth}
		\centering
		\includegraphics[width=1.1\linewidth]{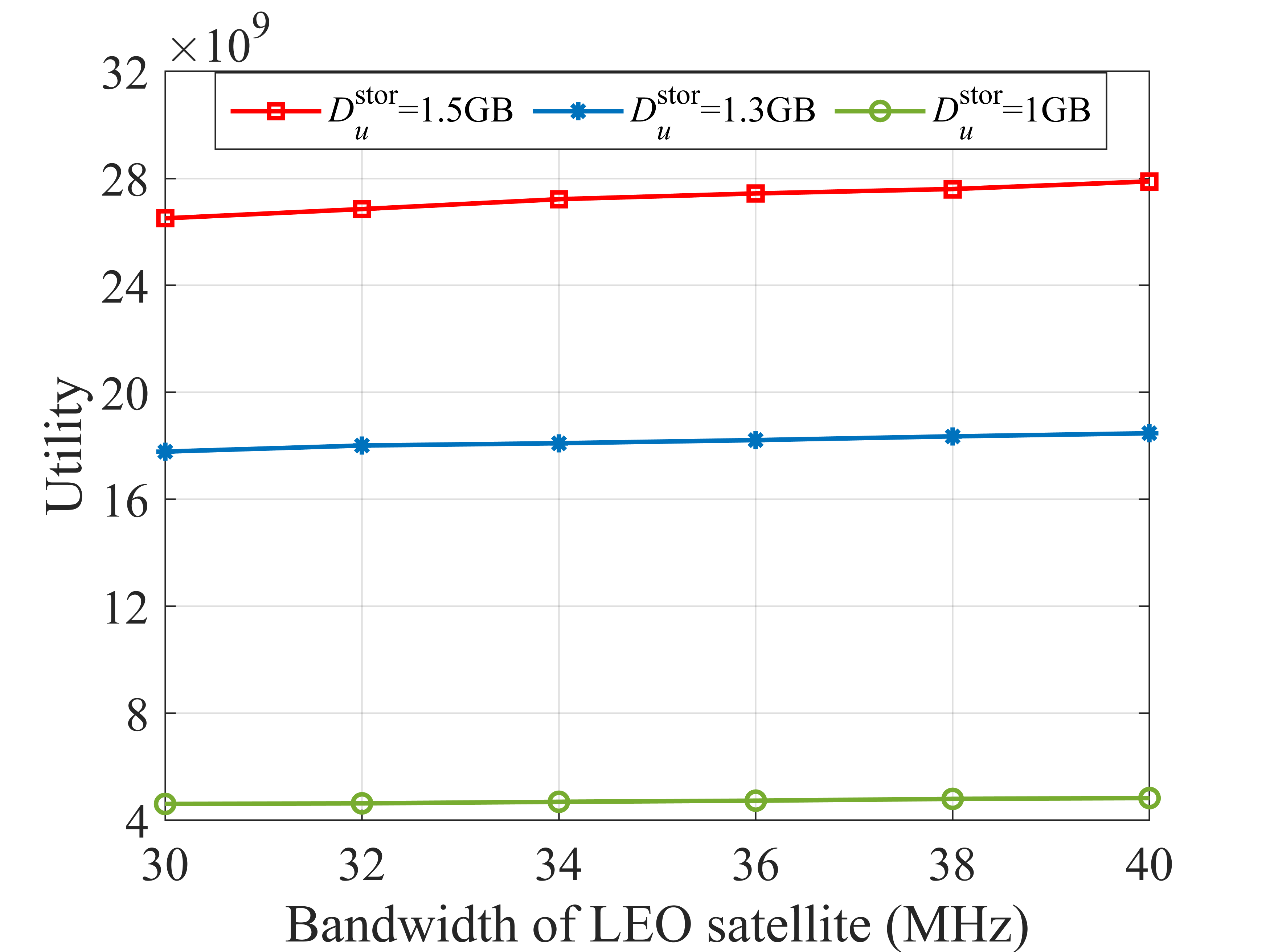}
        \centerline{(a)}
	\end{minipage}%
	\begin{minipage}[t]{0.5\linewidth}
		\centering
		\includegraphics[width=1.1\linewidth]{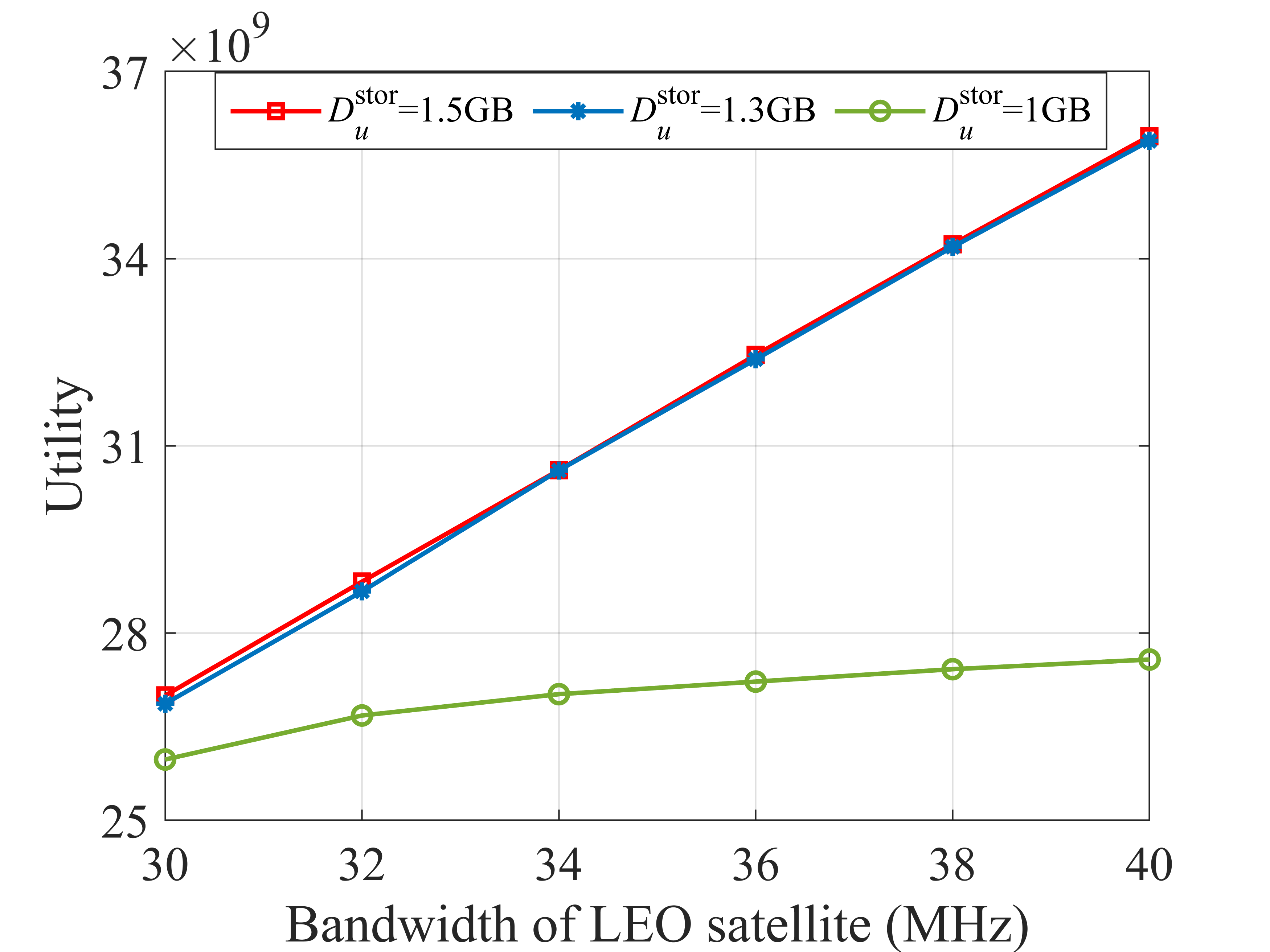}
 \centerline{(b)}
	\end{minipage}
    \vspace{6pt} 
      \caption{The relationship between utility and bandwidth of the LEO satellite under different storage space of the UAV: (a) the initial remaining storage space of the UAV is 1 GB and (b) The initial remaining storage space of the UAV is 0.5 GB.}
		\label{Fig.6}
\end{figure}

Fig. \ref{Fig.6} illustrates the relationship between the utility obtained by the JCORM algorithm and the bandwidth of the LEO satellite under different storage space for UAVs. Obviously, as storage space increases, the utility also increases. The reason is that under the condition of a constant initial remaining storage space of the UAV, the storage space of the UAV decreases and the data volume stored of the UAV in the initial condition decreases, thus reducing the data to be collected by the LEO satellite. Although an increase in the bandwidth of the LEO satellite improves its data collection capability, the total data volume is limited. As shown in Fig. \ref{Fig.6}(a), when the initial remaining storage space of the UAV is 1 GB, the increase in utility is not significant as the bandwidth increases. For instance, with a large initial storage space of 1.5 GB, an increase in LEO satellite bandwidth from 30 MHz to 40 MHz yields a modest utility gain of only 4.9\%. In contrast, when the initial remaining storage space of the UAV is reduced to 0.5 GB, the same bandwidth enhancement delivers a substantial utility increase of 24.9\%, as shown in Fig. \ref{Fig.6}(b). The reason is when the initial remaining storage space of the UAV is reduced to 0.5 GB, it indicates that there are more data stored in the UAV. During the LEO satellite data collection process, more data can be collected. However, due to the limited data collection capabilities of LEO satellite, when the storage space of the UAV is 1.3 GB or 1.5 GB, the utility difference is not significant.

\begin{figure}[t]
\centering 
{\includegraphics[width=0.35\textwidth]{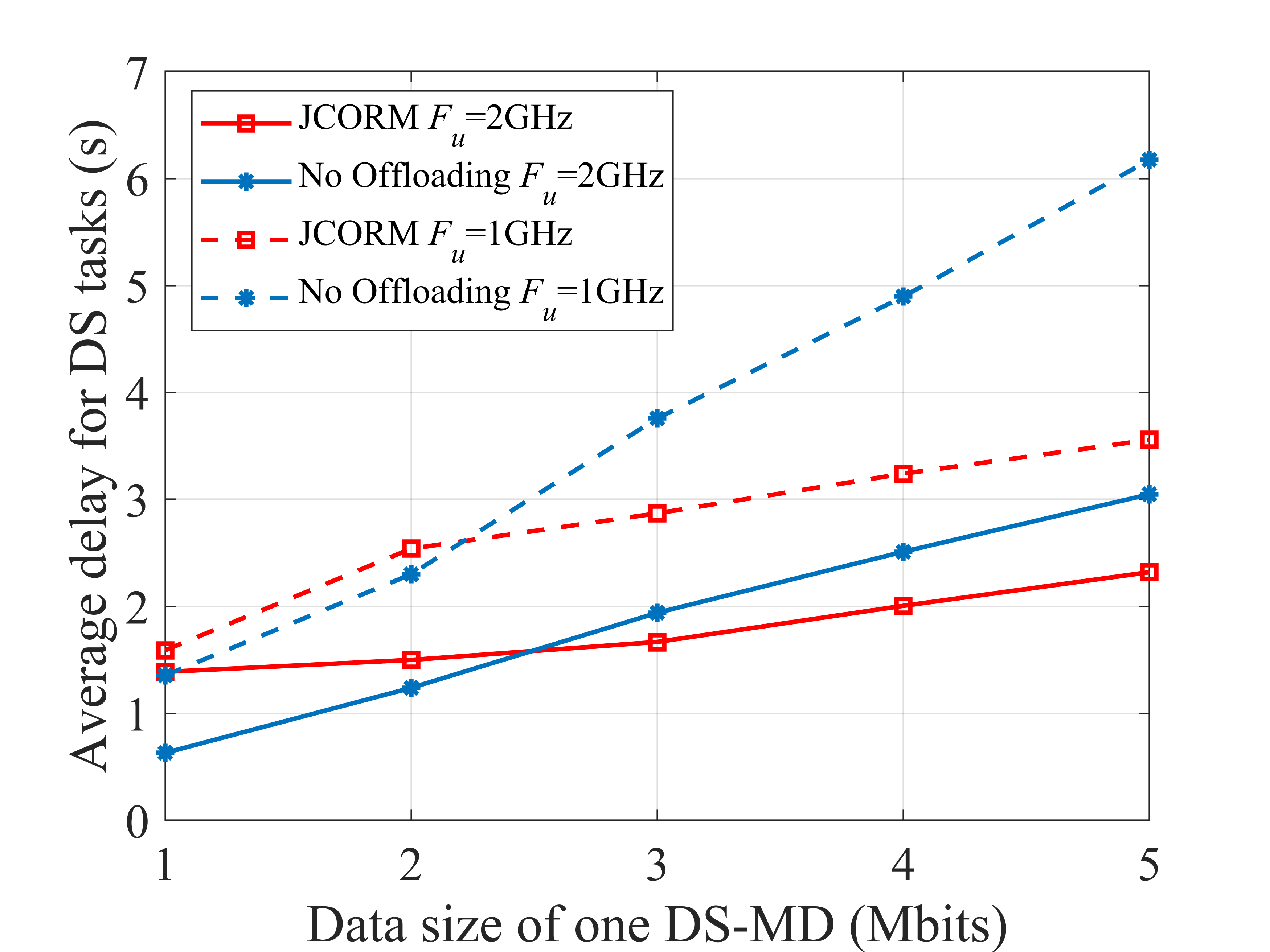}}
\caption{The relationship between the average delay for DS tasks and the data size of one DS-MD under JCORM algorithm and No Offloading scheme.}
\vspace{-3pt}
\label{Fig.7}
\end{figure}

Fig. \ref{Fig.7} shows the relationship between the average delay for DS tasks and the data size of one DS-MD under the JCORM algorithm and the No Offloading scheme. When the data size of one DS-MD is small, the computing capacity of the UAV is sufficient to handle DS tasks. The UAV tends to process tasks locally to avoid the transmission delay from the UAV to the LEO satellite, resulting in a slightly higher delay in the JCORM algorithm compared to the No Offloading scheme. However, as the data size of one DS-MD gradually increases, the average delay for DS tasks using the JCORM algorithm is significantly lower than that of the No Offloading scheme. The reason is that in the case of large data volumes, part of the DS data are offloaded to the LEO satellite for collaborative processing, which can effectively reduce the computing time. Additionally, as the computing capability of the UAV improves, the local computing delay is further reduced, and the offloading ratio decreases accordingly, thereby optimizing the efficiency of DS task processing from both computational and communication perspectives.

\begin{figure}[t]
\centering 
{\includegraphics[width=0.35\textwidth]{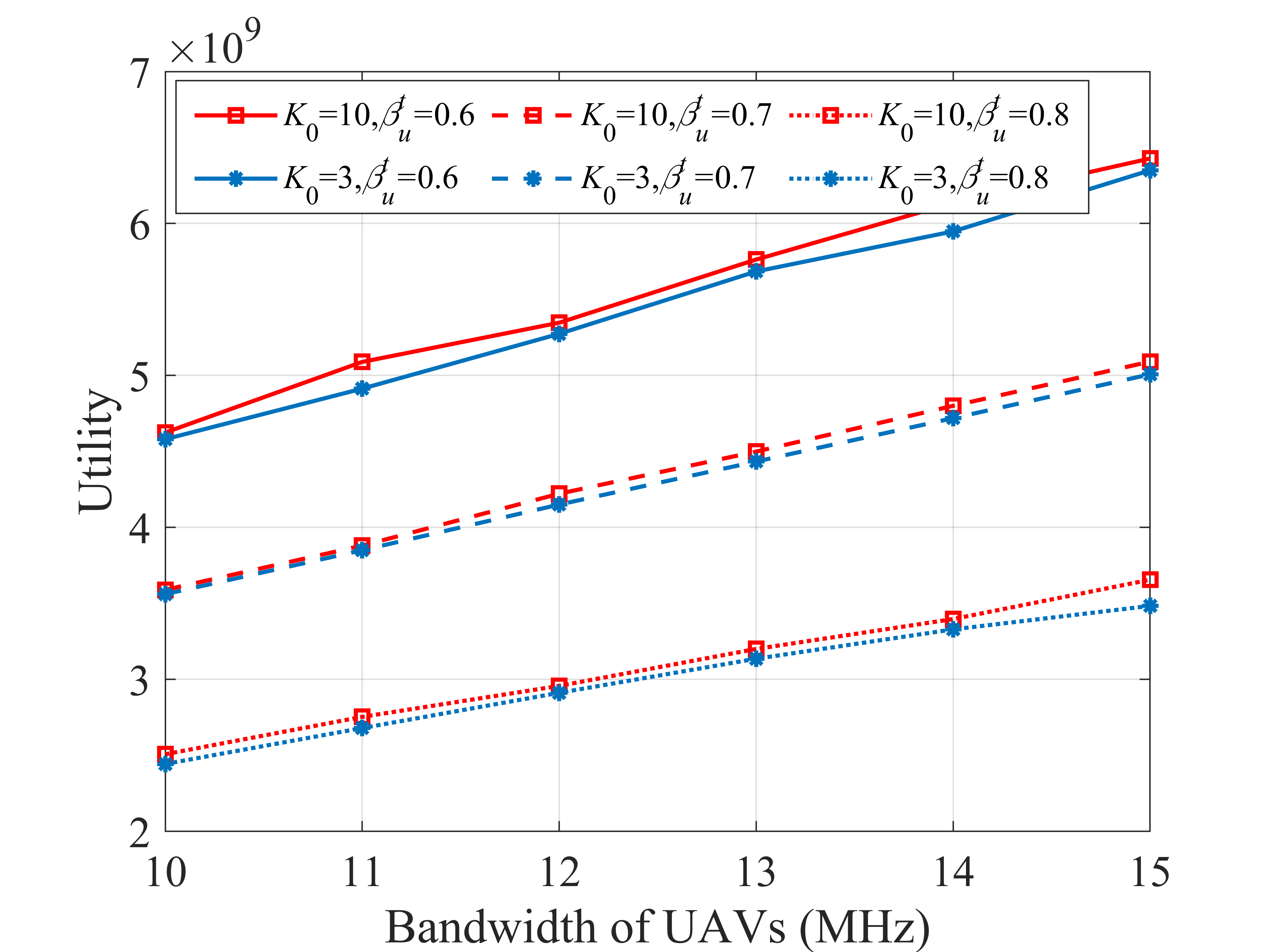}}
\caption{The relationship between the utility and the bandwidth of LEO satellite under different Rician factor.}
\vspace{-3pt}
\label{Fig.8}
\end{figure}

Fig. \ref{Fig.8} illustrates the relationship between utility and UAV bandwidth under different Rician factors. To better highlight the impact of maritime channel characteristics on system performance, we fixed both the storage capacity and the initial remaining storage space of the UAVs at 1 GB in the simulations. This setup reduces the interference from the LEO satellite collecting pre-stored data on the UAV, thereby emphasizing the effect of the maritime channel conditions on the MIoT device-to-UAV link. The Rician factor reflects the proportion of the line-of-sight component in the multi-path channel-a higher value indicates a stronger line-of-sight presence and better communication quality. As observed in the Fig. \ref{Fig.8}, for a given UAV bandwidth, the utility increases with the Rician factor $K_{\text{0}}$. This is because, in maritime environments, a higher $K_{\text{0}}$ value implies more stable and higher data transmission rate between MIoT devices and the UAV. This not only reduces the transmission time of DS data over the  MIoT device-to-UAV link but also enables the UAV to collect more DT data, collectively contributing to the growth in utility. Furthermore, when the bandwidth allocation ratio $\beta_{u}^t$increases, more bandwidth is allocated to DS tasks, resulting in a reduction in the bandwidth available for DT data collection. Consequently, under the same $K_{\text{0}}$ condition, the amount of DT data collected by the UAV decreases, resulting in a decrease in utility. 

\begin{figure}[t]
\centering 
{\includegraphics[width=0.35\textwidth]{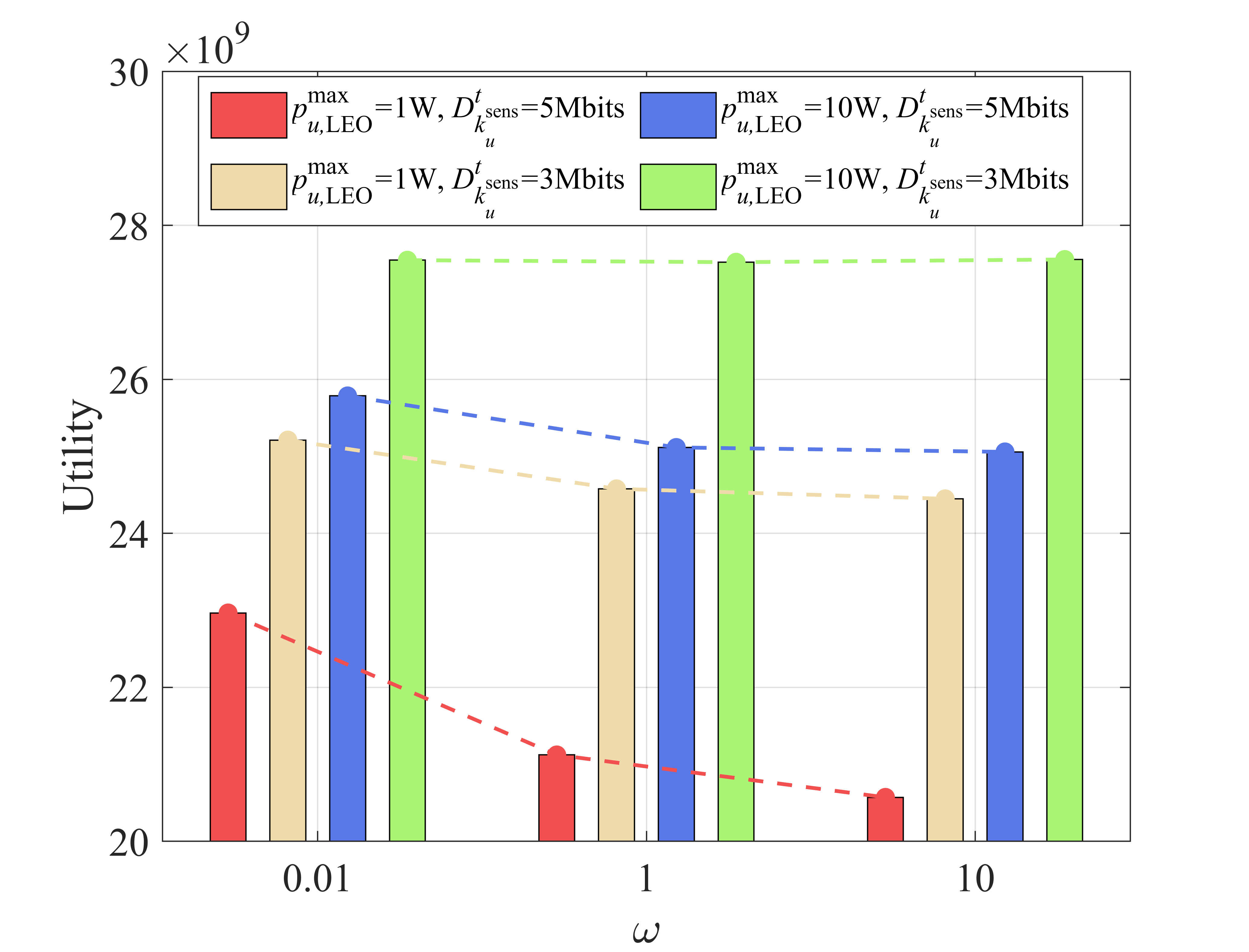}}
\caption{The relationship between the utility and $\omega$ under different maximum transmit power of the UAV and data size of one DS-MD.}
\vspace{-3pt}
\label{Fig.9}
\end{figure}

Fig. \ref{Fig.9} analyzes the impact of $\omega$ on the utility under different conditions of UAV maximum transmit power and DS-MD data size. As $\omega$ increases, the utility shows a decreasing trend. This is because $\omega$ is used to adjust the penalty assigned to energy consumption in the objective function-a larger $\omega$ implies a stronger emphasis on penalizing energy consumption, thereby leading to a reduction in utility. When $\omega$ increases from 0.01 to 1, the utility decreases noticeably. In contrast, as $\omega$ further increases from 1 to 10, the change in utility becomes more gradual. This is because the amount of data collected by the LEO satellite is generally much larger in magnitude than the total energy consumption. As a result, variations in $\omega$ within a relatively small range have a diminished influence on utility. Furthermore, under conditions of lower UAV transmit power or larger DS-MD data sizes, the utility is lower, yet the variation trend remains consistent with the above observations. This further confirms $\omega$ in balancing data collection against energy consumption in CSAMN.

\section{Conclusion}
In this paper, we have considered the problem of maximizing the utility of the CSAMN. Our approach involves optimizing the computing resource allocation, offloading ratio, transmit power of the UAV and the start time of DT tasks to achieve this goal. In particular, we have considered a scenario where heterogeneous tasks coexist, namely DS tasks and DT tasks. Furthermore, we have designed a JCORM algorithm by decomposing the optimization problem into four sub-problems and deriving the corresponding closed-form solutions. We have conducted extensive numerical simulations to validate the effectiveness of the proposed solution. The numerical results have demonstrated that proposed solution achieves more efficient utilization of network resources compared to other benchmarks.

\bibliographystyle{IEEEtran}
\bibliography{ref}


\end{document}